\documentclass{ametsoc}

\journal{jpo}

\usepackage{ulem}

\usepackage{comment}
\usepackage{enumerate}
\usepackage{bigints}

\bibpunct{(}{)}{;}{a}{}{,}

\newcommand\xoutpars[1]{\let\helpcmd\xout\parhelp#1\par\relax\relax}
\newcommand\soutpars[1]{\let\helpcmd\sout\parhelp#1\par\relax\relax}
\long\def\parhelp#1\par#2\relax{%
  \helpcmd{#1}\ifx\relax#2\else\par\parhelp#2\relax\fi%
}

\graphicspath{ {Figures/} }

\ifdraft
\else
\usepackage{doi}
\usepackage{hyperref}
\hypersetup{
	breaklinks=true,
	colorlinks=true,
	linkcolor=blue,
	citecolor=Purple,
	pdfstartview={FitH},
	pdfview={FitH 0},
	pdfauthor={N A Bakas, N C Constantinou and P J Ioannou},
	pdftitle={Statistical state dynamics of weak jets in barotropic beta-plane turbulence},
 }
\fi

\usepackage{cancel}

\usepackage{scalerel}
\usepackage[usestackEOL]{stackengine}
\let\oldmathchoice\mathchoice
\usepackage{breqn}
\let\newmathchoice\mathchoice

\def\dashint{\let\mathchoice\oldmathchoice\,\ThisStyle{\ensurestackMath{%
            \stackinset{c}{.2\LMpt}{c}{.5\LMpt}{\SavedStyle-}{%
            \SavedStyle\phantom{\int}}}%
        \setbox0=\hbox{$\SavedStyle\int\,$}\kern-\wd0}\int%
        \let\mathchoice\newmathchoice}

\newcounter{saveeqn}%

\newcommand{\bdm}{\begin{equation*}}
\newcommand{\edm}{\end{equation*}}
\newcommand{\bea}{\begin{eqnarray}}
\newcommand{\eea}{\end{eqnarray}}

\newcommand{\partialf}[2]
{
 \ifthenelse{\equal{#1}{}}{\frac{\partial}{\partial #2}}{\frac{\partial #1}{\partial #2}}
}

\newcommand{\imag}{\mathop{\mathrm{Im}}}

\renewcommand{\(}{\left(}
\renewcommand{\)}{\right)}
\renewcommand{\[}{\left[}
\renewcommand{\]}{\right]}

\newcommand{\Del}{\Delta}

\renewcommand{\d}{\delta}

\newcommand{\df}{d}

\newcommand{\s}{\sigma}

\newsavebox{\astrutbox}
\sbox{\astrutbox}{\rule[-5pt]{0pt}{20pt}}

\newcommand{\e}{\varepsilon}

\def\bit{\vphantom{\dot{W}}}

\def\thet{\vartheta}

\def\Rcal{\mathcal{R}}

\def\spb{\vphantom{\int\limits_{\thet_{j-1}+\d}^{\thet_j-\d\thet} }}

\newcommand{\defn}{\ensuremath{\stackrel{\mathrm{def}}{=}}}
\renewcommand{\equiv}{\defn}

\renewcommand{\dashint}{\int_\infty}

\newcommand{\um}{\bar{u}}

\title{A geometric interpretation of zonostrophic instability}

\authors{Georgios Kontogiannis}
 \affiliation{Laboratory of Meteorology and Climatology, Department of Physics, University of Ioannina, Ioannina, Greece}

\extraauthor{Nikolaos A.  Bakas\correspondingauthor{Nikolaos Bakas, F2-316, Department of Physics, University of Ioannina, 45110 Ioannina, Greece.}}
\extraaffil{\ifdraft\else\footnotesize\fi Laboratory of Meteorology and Climatology, Department of Physics, University of Ioannina, Ioannina, Greece}

\email{nbakas@uoi.gr}

\abstract{The zonostrophic instability that leads to the emergence of zonal jets in barotropic beta-plane turbulence is analyzed
through a geometric decomposition of the eddy stress tensor. The stress tensor is visualized by an eddy variance
ellipse whose characteristics are related to eddy properties. The tilt of the ellipse principal axis is the tilt
of the eddies with respect to the shear, the eccentricity of the ellipse is related to the eddy anisotropy, while its
size is related to the eddy kinetic energy. Changes of these characteristics are
directly related to the vorticity fluxes forcing the mean flow. The statistical state dynamics of the
turbulent flow closed at second order is employed
as it provides an analytic expression for both the zonostrophic instability and the stress tensor. For the
linear phase of the instability, the stress tensor is analytically calculated at the stability boundary. For the
non--linear equilibration of the instability the tensor is calculated in the limit of small supercriticality in which the
amplitude of the jet velocity follows Ginzburg--Landau dynamics. It is found that dependent
on the characteristics of the forcing, the jet is accelerated either because the jet primarily
anisotropizes the eddies so as to produce upgradient fluxes or because the jet changes the eddy tilt.
The instability equilibrates as
these changes are partially
reversed by the non--linear jet-eddy dynamics.}

\begin{document}

\maketitle


\section{Introduction\label{sec:intro}}

The interaction of small--scale turbulent eddies with large--scale motions in the ocean
significantly influences the ocean circulation in many ways \citep{Danabasoglu-etal-1994,Marshall-Speer-2012}.
In the Southern Ocean, the eddies undertake most of the poleward heat transfer \citep{Walkden-etal-2008}
and have an order one effect in the dynamics of the Antarctic Circumpolar Current (ACC)
\citep{Danabasoglu-etal-1994,Marshall-Speer-2012} as well as in setting the mean stratification \citep{Cessi-Fantini-2004}. In
the Western Boundary Currents, the momentum transfer by the eddies maintains
the mean jet and significantly influences its structure \citep{Waterman-Hoskins-2013} as well as produces significant jet
variability \citep{Qiu-2000}. Consequently, there is great interest in studying these interactions. The goal is twofold.
The first is understanding the physical mechanisms and dynamics setting the mean circulation, stratification,
and transport in the ocean. The second is that the ocean components of climate models do not
routinely resolve the mesoscale eddies even in the mid-latitudes where the dominant eddy scale is large. Therefore,
the effect of the eddies needs to be parameterized and future climate projections critically depend on the
skill of such parameterizations.


\cite{Marshall-etal-2012} developed a framework for studying and parameterizing the effect of the small--scale
eddies on the large scale flow, a framework that was recently termed GEOMETRIC (Geometry and Energetics of Ocean Mesoscale Eddies and
Their Rectified Impact on Climate) by David Marshall and collaborators. The core of this approach is that
the eddy forcing of the mean flow is determined by the divergence of the Eliassen–-Palm flux tensor. This tensor has a
Reynolds stress component that can be visualized by a hyperboloid and an eddy form stress tensor that can be visualized by
an ellipsoid. A more concise representation of the Reynolds stress geometry can be obtained by considering a
horizontal section of the hyperboloid resulting in the horizontal eddy variance ellipse already utilized in
barotropic studies \citep{Hoskins-etal-1983}. Similarly, a section of the ellipsoid with the vertical plane along the major axis of the
ellispoid results in a vertical eddy variance ellipse that represents the geometry of the eddy form stress tensor more compactly
\citep{Poulsen-etal-2019}. The size and the eccentricity of the ellipses are directly related to the amplitude of the eddy kinetic energy and the
eddy anisotropy respectively, while the orientation of the ellipses with respect to the background shear indicates
whether the eddies extract energy from the mean flow or surrender their energy to the mean flow.

Viewing the eddy--mean flow interactions through the Eliassen--Palm flux tensor offers three main advantages. The first
advantage is that variance ellipses can be directly obtained
from observations through principal component
analysis of the eddy velocity covariance tensor \citep{Preisendorfer-1988}. The variance ellipses have been used to
deduce eddy statistics and properties from observations \citep{Morrow-etal-1994,Trani-etal-2011}, to compare
observational and model surface eddy variability and isotropy \citep{Wilkin-Morrow-1994,Scott-etal-2008},
and to analyze eddy anisotropy and its dependence on bathymetry in ocean models \citep{Stewart-etal-2015}.

The second advantage is that the tensor can be utilized to diagnose and analyze the eddy-mean flow interactions.
\cite{Waterman-Hoskins-2013} and \cite{Waterman-Lilly-2015} considered a simplified model
of a western boundary current extension jet and calculated the statistics of the eddy geometry. They showed
that the variance ellipse patterns agreed with predictions of the jet instability and that both
the angle and anisotropy of the eddies are important for the evolution of the mean flow.
\cite{Youngs-etal-2017} utilized the GEOMETRIC framework to diagnose the role of barotropic and
baroclinic instability in the energy exchange between a standing meander and a zonal jet in a channel model
of the Antarctic Circumpolar Current and recently \cite{Poulsen-etal-2019} deduced the vertical ellipses and
the eddy form stress in the ACC from model simulations in realistic configurations.

The third advantage is that the tensor has properties that make parameterization efforts dynamically constrained
and more tractable. The reasons are first of all that parameterizations based on this tensor conserve momentum
by construction. This should be contrasted to other attempts,
as for example down–gradient closures of potential vorticity that may not satisfy momentum constraints and require
energetic constraints of the mixing efficiency to yield realistic flows \citep{Marshall-Adcroft-2010}. Additionally,
with the energy specifying the radius of the variance ellipses the
resulting tensor is written in terms of five bounded dimensionless parameters that are related to the geometry of the variance
ellipses such as their eccentricity and their orientation. The goal is then to develop parameterizations of these ellipse
characteristics and it is anticipated that the bounds on these parameters makes the parameterization efforts more tractable. While a simplified
version of such a parameterization has been recently implemented with encouraging results \citep{Mak-etal-2017,Mak-etal-2018},
further research is needed in linking
eddy--mean flow feedbacks to the characteristics of the
variance ellipses and in constraining and parameterizing the relevant parameters in simple examples.

In that vein, \cite{Marshall-etal-2012} applied the framework to the Eady model \citep{Eady-1949} and
showed that if the ellipse tilt is consistent with eddy growth, i.e leaning against the shear, then the framework
yields a correct
order of magnitude for the growth rate based on dimensional grounds only. \cite{Tamarin-etal-2016} applied
the framework to barotropic instability and obtained analytic solutions relating the eddy ellipse geometry
with the unstable modes. In this work we seek to extend these efforts to the simplified model
of forced--dissipative beta-plane turbulence.

In this model, small--scale turbulence supported by random stirring self-organizes
in large--scale structures such as zonal jets \citep{Vallis-Maltrud-93} and large scale waves that remain phase coherent
over long time scales and can even exhibit non-dispersive characteristics \citep{Galperin-etal-2010}. The jets and
the waves were shown to emerge in the flow as symmetry--breaking bifurcations \citep{Srinivasan-Young-2012,Bakas-Ioannou-2013-prl,
Constantinou-etal-2014}. That is, as the energy input rate of the forcing crosses a critical threshold value, the
flow transitions
from a homogeneous turbulent state to an inhomogeneous state with the spontaneous emergence of jets or large--scale waves.

The emergence of large--scale structures as a bifurcation, is the result of the
cooperative interaction between the small--scale turbulence and the emergent flows giving rise to a collective
type of instability \citep{Farrell-Ioannou-2007-structure}. This collective type of instability involves the
small residual of the statistical mean of the turbulent Reynolds stresses that influences the mean flow coherently,
which then modifies the statistics of the distribution of the turbulent eddies to reinforce itself. Therefore a
framework addressing the dynamics of the flow statistics is required to analytically express the instability. This
statistical state dynamics (SSD) is tractable only with a closure assumption, as a straightforward calculation leads to an
infinite hierarchy of equations for the moments \citep{Hopf-1952}. A large number of studies in the literature
on diverse physical problems ranging from quasi-geostrophic \citep{DelSole-04,Farrell-Ioannou-2008-baroclinic,
Farrell-Ioannou-2009-equatorial,Marston-2010} and stratified turbulence
\citep{Fitzgerald-Farrell-2018,Fitzgerald-Farrell-2018-jas} to turbulence in astrophysical flows \citep{Farrell-Ioannou-2009-plasmas,Tobias-etal-2011,Parker-Krommes-2013,Constantinou-Parker-2018} and in pipe
flows \citep{Constantinou-etal-Madrid-2014,Farrell-Ioannou-2017-bifur} have shown that
a second--order closure of the SSD is accurate in capturing the characteristics and dynamics of the dominant large--scale
structures. Such closures of the SSD are either referred to as stochastic
structural stability theory (S3T) \citep{Farrell-Ioannou-2003-structural} or second--order cumulant
expansion (CE2) \citep{Marston-etal-2008}.

The collective flow--forming instability was studied
within the S3T framework by  \cite{Farrell-Ioannou-2007-structure} and by \cite{Srinivasan-Young-2012}; the latter
termed the instability as zonostrophic. In these studies, the stability thresholds were calculated along with the
characteristics
of the unstable modes. \cite{Constantinou-etal-2014} and \cite{Bakas-Ioannou-2014-jfm} compared
the instability predictions to direct numerical simulations and showed that they are accurate as long
as the dynamics and effect of large--scale waves is taken into account. The equilibration of zonostrophic
instability at parameter values just above the stability threshold, was studied by
\citet{Parker-Krommes-2014-generation} who showed that the velocity amplitude of the emerging jets follows
Ginzburg--Landau (G--L) dynamics. In addition, the quantitative accuracy of the G--L approximation was examined by
comparison with jet equilibria obtained from the fully nonlinear S3T dynamics
\citep{Parker-Krommes-2014-generation,Bakas-etal-2019}.

In this work, we study the zonostrophic instability within the GEOMETRIC framework and provide a direct link
between the predictions of the S3T dynamics and the characteristics of the eddy variance ellipses. Our goal is
to elucidate the relation between the ellipse characteristics and the
eddy-mean flow feedbacks in a simple model in which the jets form and are maintained by the turbulent eddies. Jet
emergence in barotropic beta--plane turbulence presents the perfect example for this due to two
reasons. The first is that the SSD is deterministic and provides a noise-free expression of the
ellipse characteristics that in the case of jet emergence are also amenable to analytic treatment. While
the eddy-mean flow feedbacks underlying the instability and its equilibration have been previously studied
\citep{Bakas-Ioannou-2013-jas,Bakas-etal-2015,Bakas-etal-2019}, we find that treating jet formation in terms of the variance ellipses sheds
new light into the instability and its equilibration. The second is that zonostrophic
instability is markedly different from hydrodynamic
instability, in which the perturbations grow in a fixed mean flow. In the flow--forming instability, both
the coherent mean flow and the incoherent eddy field are allowed to change. The
instability manifests
as a weak zonal flow that is inserted in an otherwise homogeneous turbulent field, organizes the
incoherent fluctuations to coherently reinforce the zonal flow. Such a collective type of behavior
and support also plays a role in maintaining the large--scale flows in the atmosphere and the ocean, but has never been
treated within the GEOMETRIC framework. Here, we undertake this task and investigate
the eddy--mean flow feedbacks in this collective type of instability and further discuss possible
parameterizations of the ellipse characteristics.

This paper is organized as follows. In section 2 we present the SSD of barotropic beta-plane
turbulence as well as the zonostrophic instability that forms zonal jets in the turbulent flow.
In section 3 we review the GEOMETRIC framework for barotropic flows and discuss how changes in
the characteristics of the eddy variance ellipse force the zonal flow. In section 4 we
calculate the variance ellipse for the unstable jets and relate the change of its characteristics
to the upgradient fluxes. Finally,
in section 5 we calculate the variance ellipse during the evolution of the instability in the limit
of small supercriticality and discuss the eddy--mean flow feedbacks underlying its equilibration. We
end with the conclusions in section 6.

\section{Statistical state dynamics of barotropic $\beta$-plane turbulence and zonostrophic instability\label{sec:SSD}}

Consider a forced-dissipative barotropic flow on a $\beta$-plane. The dynamics is governed by the non-linear
equation for the evolution of relative vorticity $\tilde{\zeta}$:
\begin{equation}
\partial_{\tilde t}\tilde{\zeta}+J(\tilde{\psi},\tilde{\zeta})+\tilde{\beta}\partial_{\tilde{x}}\tilde{\psi}=
-r\tilde{\zeta}+\sqrt{\tilde{\epsilon}}\tilde{\xi},\label{eq:NL}
\end{equation}
where $\tilde{\beta}$ is the planetary vorticity gradient, $\tilde{\psi}$ is the streamfunction that is linearly related to vorticity
through the inverse of the Laplacian $\Delta=\partial_{\tilde{x}}^2+\partial_{\tilde{y}}^2$ ($\tilde{\psi}=\Delta^{-1}\tilde{\zeta}$) and $J(f,g)=\partial_{\tilde{x}}f\partial_{\tilde{y}}g-\partial_{\tilde{y}}f\partial_{\tilde{x}}g$ is the Jacobian. Bottom drag is parameterized
through linear dissipation of vorticity at a rate $r$. The random stirring $\tilde{\xi}$ parameterizes processes such
as small--scale convection or baroclinic instability and maintains the turbulence in the flow. It is assumed
to be temporally uncorrelated and spatially homogeneous with a prescribed spatial correlation function $Q$ and
to inject energy at a rate $\tilde{\epsilon}$ in the flow. That is, the
correlation of $\tilde{\xi}$ between two different points in space ($\tilde{\bf x}_a$, $\tilde{\bf x}_b$) and two different
points in time ($\tilde{t}_a$, $\tilde{t}_b$) is:
\begin{equation}
\left<\tilde{\xi}(\tilde{\bf x}_a,\tilde{t}_a)\tilde{\xi}(\tilde{\bf x}_b,\tilde{t}_b)\right>=
Q(\tilde{\bf x}_a-\tilde{\bf x}_b)\delta(\tilde{t}_a-\tilde{t}_b),
\end{equation}
where the brackets denote an average over the realizations of the forcing. In this work we consider an excitation
that injects energy in a delta ring in wavenumber space with radius
$k_f$. That is, the power spectrum of the correlation function is:
\begin{eqnarray}
\hat{Q}(\tilde{\bf k})&=&\frac{1}{2\pi}\int Q(\tilde{\bf x}_a-\tilde{\bf x}_b)
e^{-i\tilde{\bf k}\cdot (\tilde{\bf x}_a-\tilde{\bf x}_b)}d^2\tilde{\bf x}\nonumber\\
&=&2\pi k_f\delta(\tilde{k}-k_f)\left[1 + \eta\cos (2 \phi )\right],\label{eq:Qf}
\end{eqnarray}
where $\tilde{k}$ is the amplitude of the wavevector $\tilde{\bf k}=(\tilde{k}_x,\tilde{k}_y)$ and
$\phi=\arctan(\tilde{k}_y/\tilde{k}_x)$. We consider two cases of forcing that have been
used in previous studies of $\beta$-plane turbulence. An isotropic
forcing ($\eta=0$) that is thought as parametrizing vorticity sources such as convection
\citep{Lilly-1969,Vallis-Maltrud-93,Galperin-etal-2010} and an anisotropic
forcing ($\eta=1$) that injects more power in waves with small
$|\tilde{k}_y|$, as if the vorticity injection was due to baroclinic growth processes
\citep{Srinivasan-Young-2014,Bakas-etal-2015}. We non-dimensionalize equation (\ref{eq:NL}) using the forcing length scale $k_f^{-1}$ as the unit of
length and the dissipation time scale $r^{-1}$ as the unit of time. The vorticity, the streamfunction, the planetary
vorticity gradient and the energy input rate are non-dimensionalized by
$\zeta=\tilde{\zeta}/r$, $\psi=\tilde{\psi}r/k_f^2$, $\beta=\tilde{\beta}/k_{f}r$ and
$\epsilon=\tilde{\epsilon}k_f^2/r^3$ respectively. Therefore the non-dimensional versions
of (\ref{eq:NL}) and (\ref{eq:Qf}) lack the tildes and have $r=1$ and $k_f=1$.

To investigate the eddy-mean flow feedback in the zonostrophic instability and to obtain
a clear view of the eddy variance ellipse, we formulate the equations that evolve the flow statistics (SSD). We briefly discuss the
derivation of the SSD for the barotropic dynamics, which can be found in previous studies like \cite{Farrell-Ioannou-2003-structural}
and \cite{Srinivasan-Young-2012}. Typically one separates the vorticity field into a zonal mean, denoted by the overbar, and a non-zonal
eddy component, denoted by primes:
\begin{equation}
\zeta=\overline{\zeta}+\zeta'.\label{eq:sep}
\end{equation}
Substituting (\ref{eq:sep}) in the non-dimensional form of (\ref{eq:NL}), we readily obtain the equations for
the evolution of the zonal jet and the
non-zonal eddies:
\begin{eqnarray}
\partial_t\overline{u}&=&\overline{v'\zeta'}-\overline{u},\label{eq:uevo}\\
\partial_t\zeta'&=&\mathcal{A}\zeta'+\underbrace{\overline{J(\psi',\zeta')}-J(\psi',\zeta')}_{eddy-eddy}+\sqrt{\varepsilon}\xi,\label{eq:zetaevo}
\end{eqnarray}
where the zonal mean zonal velocity $\overline{u}$ is related to the mean vorticity by $\overline{\zeta}=-\partial_y\overline{u}$ and
\begin{equation}
\mathcal{A}=-\overline{u}\partial_x-(\beta-\partial_{y}^2\overline{u})\partial_x\Delta^{-1}-1,\label{eq:op_A}
\end{equation}
is the operator governing the quasi-linear jet-eddy interaction. Under a quasi-linear assumption in which the
eddy-eddy interactions noted in (\ref{eq:zetaevo}) are ignored, the SSD of (\ref{eq:uevo})-(\ref{eq:zetaevo}) closes at second order and
comprises of equations for the evolution of the first cumulant $\overline{u}(y,t)$, which is the zonal jet velocity and
the second cumulant $C({\bf x}_a, {\bf x}_b, t)=\overline{\zeta'({\bf x}_a,t) \zeta'({\bf x}_b,t)}$,
which is the eddy vorticity correlation. The equations governing the evolution of the two cumulants are
\footnote{Details on the derivation of the SSD can be found in previous studies, e.g.
\cite{Farrell-Ioannou-2003-structural,Srinivasan-Young-2012,Bakas-etal-2015}.}:
\begin{eqnarray}
\partial_t\overline{u}&=&\mathcal{R}(C)-\overline{u},\label{eq:cum1}\\
\partial_tC&=&\left(\mathcal{A}_a+\mathcal{A}_b\right)C+\varepsilon Q,\label{eq:cum2}
\end{eqnarray}
where the linear operator
\begin{equation}
\Rcal(C)=\frac{1}{2}\[\(\partial_{x_a}\Del_a^{-1}+\partial_{x_b}\Del_b^{-1}\)C\bit\]_{a=b}\ ,\label{eq:rs}
\end{equation}
gives the vorticity flux as a function of the eddy vorticity correlation $C$. The subscripts in the operators denote
differentiation with respect to the different points ${\bf x}_a$ and ${\bf x}_b$ as well as evaluation of the mean velocity
at these points. In (\ref{eq:rs}), the subscript $a=b$ denotes the evaluation of the expression in the brackets at the same
point. The closed deterministic system (\ref{eq:cum1})-(\ref{eq:cum2}), termed as S3T dynamics, comprises the joint evolution
of the jet and its associated eddy statistics.

For all values of $\beta$ and $\varepsilon$, the state with zero mean flow and a homogeneous covariance proportional
to the forcing covariance:
	\begin{equation}
		\um_0=0\ \ ,\ \ C_0({\bf x}_a-{\bf x}_b)=\frac{\epsilon}{2}Q({\bf x}_a-{\bf x}_b)\ ,\label{eq:hom_eq}
	\end{equation}
is a fixed point of the S3T dynamics. Jets emerge in the flow as this homogeneous equilibrium becomes
unstable when the energy input rate passes a critical value \citep{Farrell-Ioannou-2007-structure}. The incipient instability that is termed
zonostrophic, is addressed by linearizing (\ref{eq:cum1})-(\ref{eq:cum2}) around equilibrium (\ref{eq:hom_eq}). The eigenfunctions
consist of a harmonic jet $\delta\overline{u}=e^{iny}$ with wavenumber $n$ and a corresponding eddy covariance
that is given by:
\begin{equation}
\delta  C=\left[G^+({\bf x}_a-{\bf x}_b)-G^-({\bf x}_a-{\bf x}_b)\right] e^{in(y_a+y_b)/2}\ ,\label{eq:eigfunc}
\end{equation}
where
\begin{equation}
G^\pm({\bf x})=\int\frac{d^2{\bf k}}{2\pi}\frac{ik_xk_{\mp}^2(k_{\pm}^2-n^2)}{(\sigma+2)k_+^2k_-^2+2i\beta nk_x k_y}\frac{\hat{Q}({\bf k}_{\pm})}{2}
e^{i{\bf k}\cdot {\bf x}},\label{eq:G}
\end{equation}
${\bf n} = (0, n)$, ${\bf k}_{\pm}={\bf k}+{\bf n}/2$ and $k_\pm=|{\bf k}_{\pm}|$. The dispersion relation for
the eigenvalues $\sigma$ was shown by \cite{Bakas-etal-2015} to be:
	\begin{equation}
	  \sigma=\epsilon f(\sigma|\delta\overline{u}, Q/2)-1\ .\label{eq:dispersion}
	\end{equation}
The term $f$ is given in (\ref{eq:ff}) and
expresses the vorticity flux induced by the covariance perturbation $\delta C$. The vorticity flux $f$ as
well as the eigenvalues are real in this case for jet perturbations with scales larger than the forcing scale ($n<1$). A jet with
wavenumber $n$ is therefore rendered unstable under two conditions: (i) the vorticity fluxes $f$ are positive, that is the
distortion of the eddies by the jet as expressed by $\delta C$ produces upgradient fluxes and (ii) the amplitude of the
fluxes should be large enough to overcome dissipation. This second condition is satisfied when the energy
input rate is above a critical value
$\epsilon_t(n)$ that can be computed
from (\ref{eq:dispersion}) by assuming criticality ($\sigma=0$): $\epsilon_t=1/f(0|\delta\overline{u}, Q/2)$. The minimum
value of $\varepsilon_t$ over all wavenumbers
\begin{equation}
	\epsilon_c=\min_{ n }[\epsilon_t(n)],\label{eq:stab_bound}
\end{equation}
is the stability boundary above which jets form. The critical rate $\epsilon_c$ is
attained by the most unstable jet with wavenumber $n_c$. Details on the characteristics of the dispersion
relation of the zonostrophic instability and the stability boundary (\ref{eq:stab_bound}) can be found in
\cite{Srinivasan-Young-2012} and \cite{Bakas-etal-2015}. The goal in this work is to investigate the eddy fluxes
producing the instability through analysis of the eddy variance ellipse and to illuminate the jet-eddy interactions
leading to jet formation.

\section{The GEOMETRIC approach and the eddy variance ellipse}

In this section we review the GEOMETRIC approach as is applied for barotropic flows and closely follow
the discussion and notation in \cite{Tamarin-etal-2016}. The GEOMETRIC approach relies in expressing the vorticity fluxes that
force the mean flow in terms of the eddy
momentum stress tensor ${\bf T}$ through the Taylor identity:
\begin{equation}
\overline{{\bf u}'\zeta'}=\nabla\cdot{\bf T}, \label{eq:vor_flux}
\end{equation}
where ${\bf T}$ is for the barotropic dynamics considered \citep{Tamarin-etal-2016}:
\begin{equation}
{\bf T}=\left(
\begin{array}{ccc}
  N  &  M \\
  M & -N
\end{array}
\right),
\end{equation}
with $N=\overline{u'v'}$ the eddy momentum fluxes and $M=(1/2)\left(\overline{{v'}^2}-\overline{{u'}^2}\right)$ a quantity measuring the anisotropy
of the eddies. One of the advantages of expressing the vorticity fluxes in terms of the tensor ${\bf T}$ is that the stress tensor can be visualized
geometrically through an eddy variance ellipse, the characteristics of which (tilt, eccentricity and focal point) can be obtained at each point in the
flow. Therefore, by examining how the ellipse changes throughout the flow we can deduce, through expression (\ref{eq:vor_flux}),
the mean flow forcing.

\begin{figure}
\centerline{\includegraphics[width=35pc]{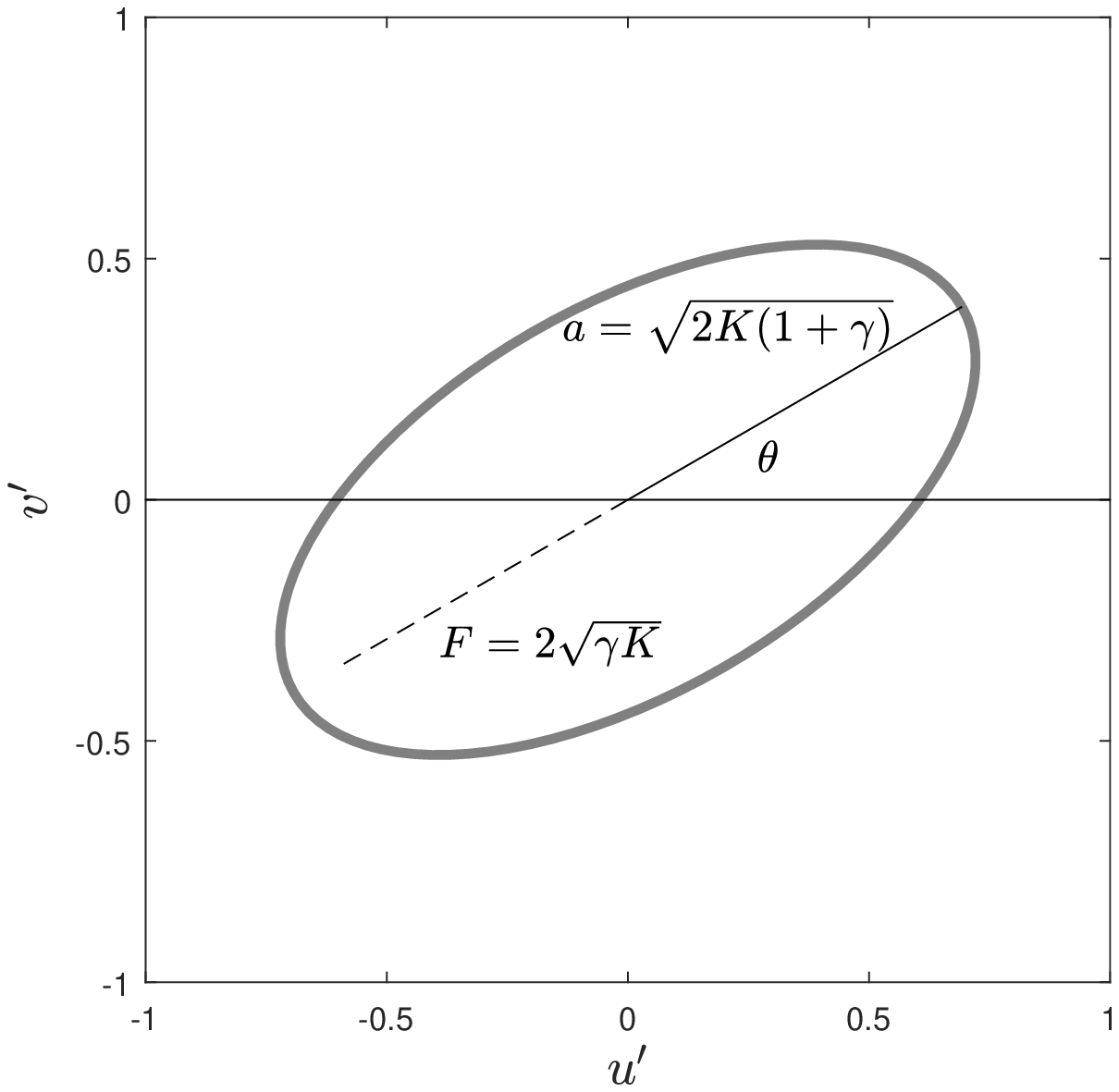}}
\caption{\label{fig:eddy_ellipse} Schematic representation of the eddy variance ellipse in the eddy velocity space. Noted are the
angle of the tilt of the ellipse $\theta$, the semi-major axis $a$ (solid), the linear eccentricity $F$ (dashed), and their expressions in terms of the anisotropy
parameter $\gamma$ and the eddy kinetic energy $K$.}
\end{figure}

The geometric representation of the ellipse becomes evident be rewriting the tensor as \citep{Marshall-etal-2012,Tamarin-etal-2016}:
\begin{equation}
{\bf T}=\gamma K\left(
\begin{array}{ccc}
   \sin{2\theta}  & -\cos{2\theta} \\
  -\cos{2\theta} & -\sin{2\theta}
\end{array}
\right), \label{eq:eddy tensor}
\end{equation}
where $K=(1/2)(\overline{{u'}^2}+\overline{{v'}^2})$ is the eddy kinetic energy, $\theta=(1/2)\arctan{(-N/M)}$ is the tilt of the
ellipse principal axis and $\gamma=\sqrt{N^{2}+M^{2}}/K$ is a non-dimensional parameter related to the
eddy anisotropy that was shown by \cite{Marshall-etal-2012} to be bounded ($0\leq\gamma\leq 1$). The eccentricity $e$ of the ellipse is determined
by the anisotropy parameter as $e=\sqrt{2\gamma/(1+\gamma)}$. The size of the ellipse is determined by
the kinetic energy and the anisotropy parameter, as the linear eccentricity (the distance $F$ between a focal point and the center of the ellipse) is $F=2\sqrt{\gamma K}$. Therefore, the
semi-major axis of the ellipse is $a=F/e=\sqrt{2K(1+\gamma)}$. The ellipse in the eddy velocity space and its characteristics are
visualized in Fig. \ref{fig:eddy_ellipse}.

\begin{figure}
\centerline{\includegraphics[width=35pc]{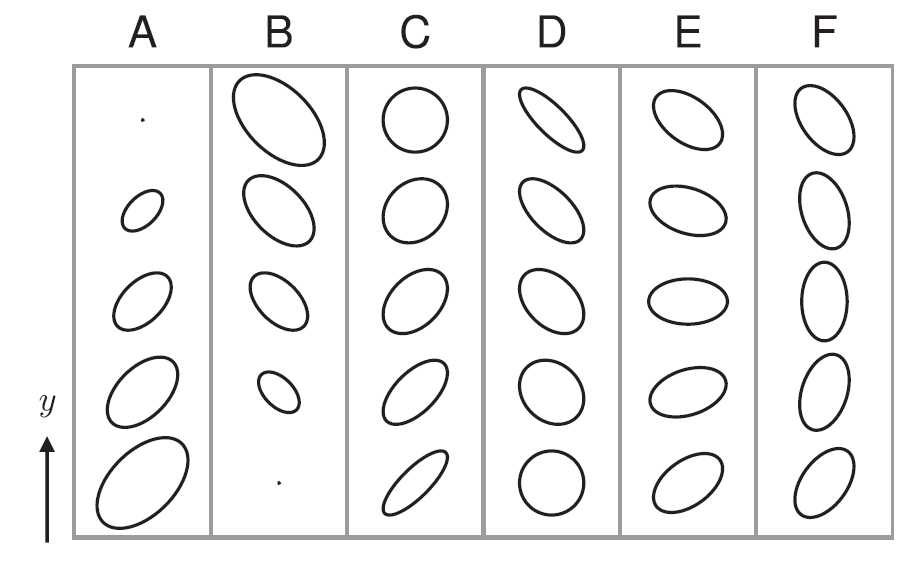}}
\caption{\label{fig:ellipse-eastward-westward} Schematic representation of the changes in the eddy variance ellipse with latitude that
lead to an eastward acceleration in the flow. In each of the columns one of the parameters $\gamma$, $\theta$ and $K$ are changing.
Column A: $K$ decreasing with $y$ and $\theta$ is in the first (or third) quadrant. Column B: $K$ increasing with $y$ and $\theta$ is in
the second (or fourth) quadrant. Column C: $\gamma$ decreasing with $y$ and $\theta$ is in the first (or third) quadrant. Column D:
$\gamma$ increasing with $y$ and $\theta$ is in the second (or fourth) quadrant. Column E: $\theta$ decreasing with $y$ and $|\theta|<\pi/4$.
Column F: $\theta$ increasing with $y$ and $|\theta|<\pi/4$. Figure taken from \cite{Tamarin-etal-2016}.}
\end{figure}

In order to relate the eddy fluxes that force the mean flow to the spatial changes in the ellipse characteristics $(\gamma,~\theta,~K)$ we substitute ${\bf T}$ from
(\ref{eq:eddy tensor}) into Taylor's identity (\ref{eq:vor_flux}) and obtain:
\begin{equation}
\overline{v'\zeta'}=-\left(\gamma\frac{\partial K}{\partial y}+K\frac{\partial\gamma}{\partial y}\right)\sin 2\theta-2\gamma K\frac{\partial\theta}{\partial y}\cos 2\theta. \label{eq:vor_flux2}
\end{equation}
Therefore, the meridional gradients of the three parameters (the three terms in equation (\ref{eq:vor_flux2}))
can lead to a jet acceleration or deceleration depending on both the sign of the gradient and the orientation of the
ellipse. Figure \ref{fig:ellipse-eastward-westward} shows the different cases that illustrate how the three
gradients influence the sign of the vorticity fluxes.
For example, we see that if the eddy kinetic energy (column B) or if the eccentricity (column D) increase with latitude and the ellipse tilt angle is in the
second (or fourth) quadrant, then this leads to an eastward acceleration of the zonal flow. An eastward acceleration also occurs if, for example, the axial tilt is $|\theta|\leq\pi/4$ and decreases with latitude (column E). Note that opposite changes to the ones shown in the columns of Fig. \ref{fig:ellipse-eastward-westward}, imply
westward acceleration. Our goal is to investigate the eddy variance ellipse characteristics at the onset of zonostrophic instability
and the contribution
of the changes of the three characteristics to the positive vorticity flux feedback that underlies the instability.

\section{The eddy variance ellipse at the onset of zonostrophic instability}

\begin{figure*}
\centerline{\includegraphics[width=35pc]{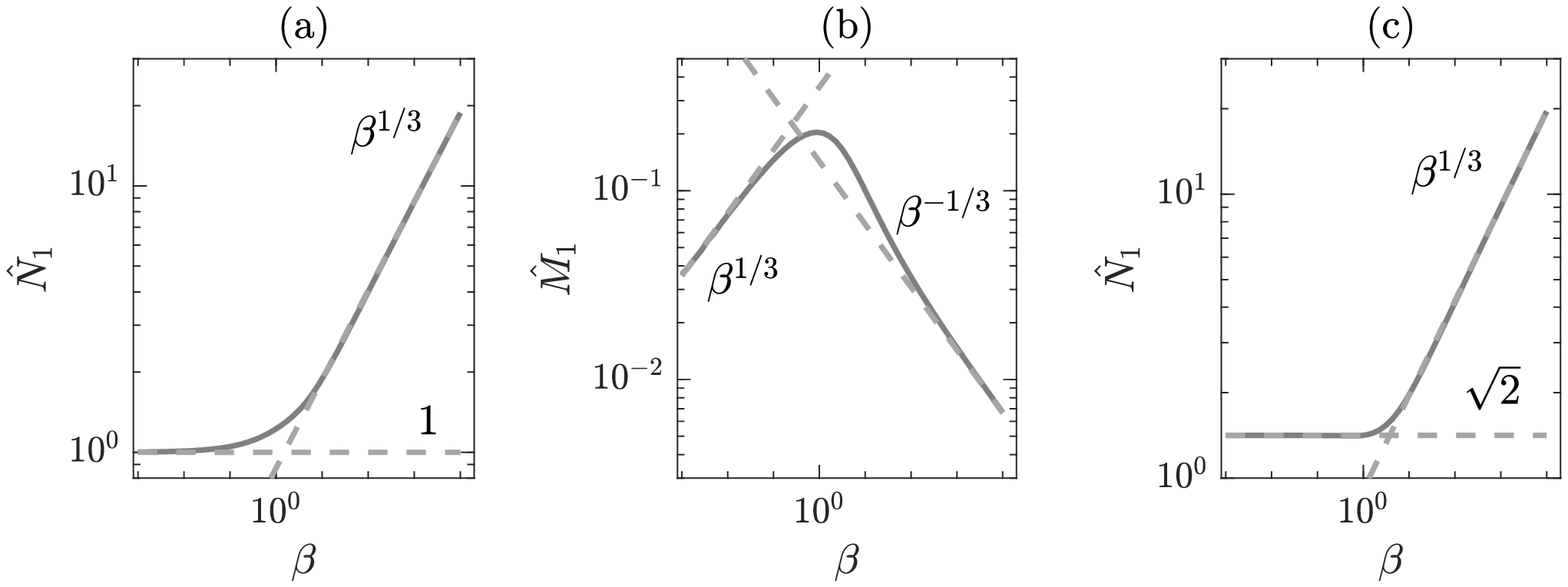}}
\caption{\label{fig:fluxes} (a)-(b) Amplitude of (a) the momentum fluxes $\hat{N}_1$ and (b) the eddy anisotropy
$\hat{M}_1$ calculated at the onset of zonostrophic instability as a function of $\beta$ for isotropic forcing ($\eta=0$). (c) Amplitude of the momentum
fluxes $\hat{N}_1$ calculated at the onset of zonostrophic instability as a function of $\beta$ for anisotropic forcing ($\eta=1$).
Also shown are the asymptotic expressions (\ref{eq:Nbs}), (\ref{eq:Mbs}) and (\ref{eq:Mbl}) (dashed lines).}
\end{figure*}

To obtain the eddy variance ellipse at the onset of zonostrophic instability, we calculate
the eddy momentum fluxes $N$, the eddy anisotropy $M$ as well as the eddy kinetic energy $K$ for the most
unstable jet $\overline{u}_1=\sin(n_cy)$ with wavenumber $n_c$ at onset ($\sigma=0$):
\begin{eqnarray}
N & = & N_1=\hat{N}_1(\beta)\cos(n_cy), \label{eq:N_dn} \\
M & = & M_0+M_1=M_0 +\hat{M}_1(\beta)\sin(n_cy), \label{eq:M_dm} \\
K & = & K_0+K_1=K_0+\hat{K}_1(\beta)\sin(n_cy). \label{eq:K_dk}
\end{eqnarray}
The subscript $0$ denotes the quantities calculated for the homogeneous equilibrium (\ref{eq:hom_eq}).
The subscript $1$ denotes the small
perturbations induced by the jet $\overline{u}_1$ and
whose amplitudes depend on the planetary vorticity gradient $\beta$. Note
that for the homogeneous equilibrium the momentum fluxes are zero.
We first calculate the stress tensor for the isotropic forcing ($\eta=0$). In Appendix A we show that in this case, $M_0=0$ due to the isotropy of
the forced eddies, $K_0=\epsilon_c /2$, and $\hat{N}_1$, $\hat{M}_1$, are given by the integrals (\ref{eq:mf}), (\ref{eq:eddyan}).
Asymptotic values of the integrals in the limits of small and large $\beta$ are also derived (Eqs. (\ref{eq:Nbs}), (\ref{eq:Mbs})
and (\ref{eq:Mbl})). Figure \ref{fig:fluxes} shows the amplitudes $\hat{N}_1$ and $\hat{M}_1$ as a function of $\beta$. We observe that
the amplitude of the momentum fluxes $\hat{N}_1$ is a monotonically increasing function of $\beta$ reaching the minimum value of
$\hat{N}_1\simeq 1$ for $\beta\ll1$ while increasing as $\hat{N}_1\simeq
\left(2\beta/3\right)^{1/3}$ for $\beta\gg1$. The amplitude of the eddy anisotropy $\hat{M}_1$ asymptotically increases as
$\hat{M}_1\simeq \left(12\beta/270\right)^{1/3}$ for $\beta\ll1$, it reaches a maximum value for $\beta\sim\mathcal{O}(1)$ and subsequently
decreases as $\hat{M}_1\simeq \left(3/2^{10}\beta\right)^{1/3}$ for $\beta\gg1$.

\begin{figure*}
\centerline{\includegraphics[width=35pc]{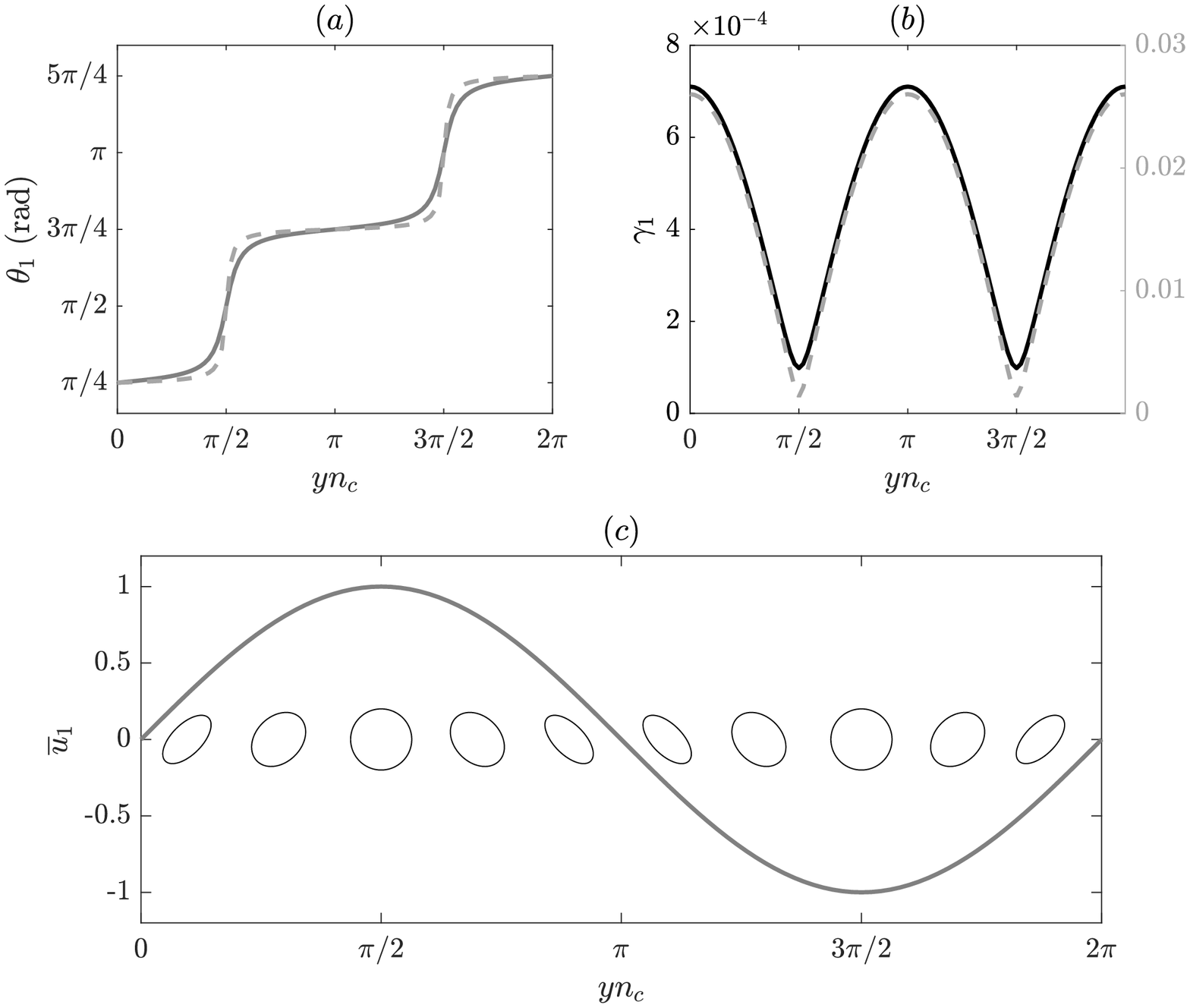}}
\caption{\label{fig:ellipse_change} (a) Tilt angle $\theta_1$ and (b) anisotropy parameter $\gamma_1$ as a function of latitude
for the most unstable jet perturbation $\overline{u}_1=\sin(n_cy)$ and $\beta=0.1$ (solid line) and $\beta=10$ (dashed line). (c)
Schematic change of the variance ellipse within the most unstable jet perturbation
shown by the thick line. The ellipse
eccentricity is of the same order as the infinitesimal amplitude of the mode ($\mathcal{O}(\mu)\ll1$), so changes in the ellipse
eccentricity are exaggerated by $1/\mu$ for illustration purposes. The forcing is isotropic ($\eta=0$).}
\end{figure*}

The variance ellipse characteristics can then be readily derived. The tilt and the anisotropy
parameter are to leading order:
\begin{equation}
\theta_1=\frac{1}{2}\arctan\left(-\frac{\hat{N}_1}{\hat{M}_1}\cot(n_cy)\right),
\end{equation}
\begin{equation}
\gamma_1=\frac{\sqrt{\hat{N}_1^{2}\cos^2(n_cy)+\hat{M}_1^{2}\sin^2(n_cy)}}{K_0}.
\end{equation}
Figure \ref{fig:ellipse_change} shows the tilt and the anisotropy parameter as a function of latitude for
a wavelength of the most unstable jet and for two values of $\beta$. The tilt angle for
both values of $\beta$ is piecewise constant assuming the values of $\pi/4$ in the first quarter of the
jet wavelength, $3\pi/4$ in the second and third quarter of the jet wavelength and $5\pi/4$ in the last quarter. The angle increases within a
very narrow width of latitudes in the vicinity of the jet cores ($yn_c\simeq \pi/2$ and $yn_c\simeq 3\pi/2$). The reason for
the piecewise constant profile of the tilt is that based on the asymptotic scalings (\ref{eq:Nbs}) and (\ref{eq:Mbs}), the
amplitude of the eddy anisotropy is much smaller than the amplitude of the momentum fluxes ($\hat{M}_1\ll \hat{N}_1$) for both low
and high values of $\beta$.
As a result their ratio is large yielding values that are odd multiples of $\pi/4$ for the angle $\theta_1$. The only latitudes over which the tilt is
different, are near the jet core ($|yn_c|\sim \pi/2$) where the momentum fluxes are small as well and
$\hat{N}_1\cot(n_cy)/\hat{M}_1\sim\mathcal{O}(1)$. The reason is that symmetry considerations require an angle of zero or
$\pi/2$ right at the jet core \citep{Tamarin-etal-2016}. In contrast, we have
significant changes in the anisotropy parameter $\gamma_1$ over all latitudes. Since the amplitude of the eddy
anisotropy ($\hat{M}_1$) is much smaller than the amplitude of the momentum fluxes ($\hat{N}_1$)
the anisotropy parameter $\gamma_1\propto |\cos(n_cy)|$. Therefore the eddies are
almost isotropic at the jet core where we have minimum values of $\gamma_1$ and anisotropic at the regions of
maximum shear where we have maximum values of $\gamma_1$.

The change of these parameters leads to concomitant changes of the variance ellipse with latitude that are schematically shown in
Fig. \ref{fig:ellipse_change}(c). First of all, we observe that the eddies lean with the jet shear, which is consistent
with the intuitive picture of the eddies surrendering their energy to the mean flow. In the first quarter of the jet wavelength, the
increase of the tilt and the decrease of the anisotropy parameter
with latitude corresponds to patterns C and F of Fig. \ref{fig:ellipse-eastward-westward}, leading to an eastward acceleration
of the flow. An eastward acceleration is induced in the second quarter of the wavelength as well, due to the increase of the tilt angle and
the anisotropy parameter with latitude with $|\theta|>\pi/4$ (patterns D and F). In the
second half of the jet wavelength, the increase of tilt with latitude corresponds to the opposite pattern of D and leads to a westward
acceleration. Similarly, the increase of the ellipse eccentricity in the third quarter of the wavelength with
the tilt angle in the second quadrant and the decrease of the ellipse eccentricity in the last quarter of the jet wavelength with the
tilt angle in the third quadrant correspond to the opposite patterns of D and C and lead to westward acceleration as well. As a result, both the
changes in the tilt and the eccentricity reinforce the jet perturbation.

\begin{figure}
\centering\includegraphics[width=35pc]{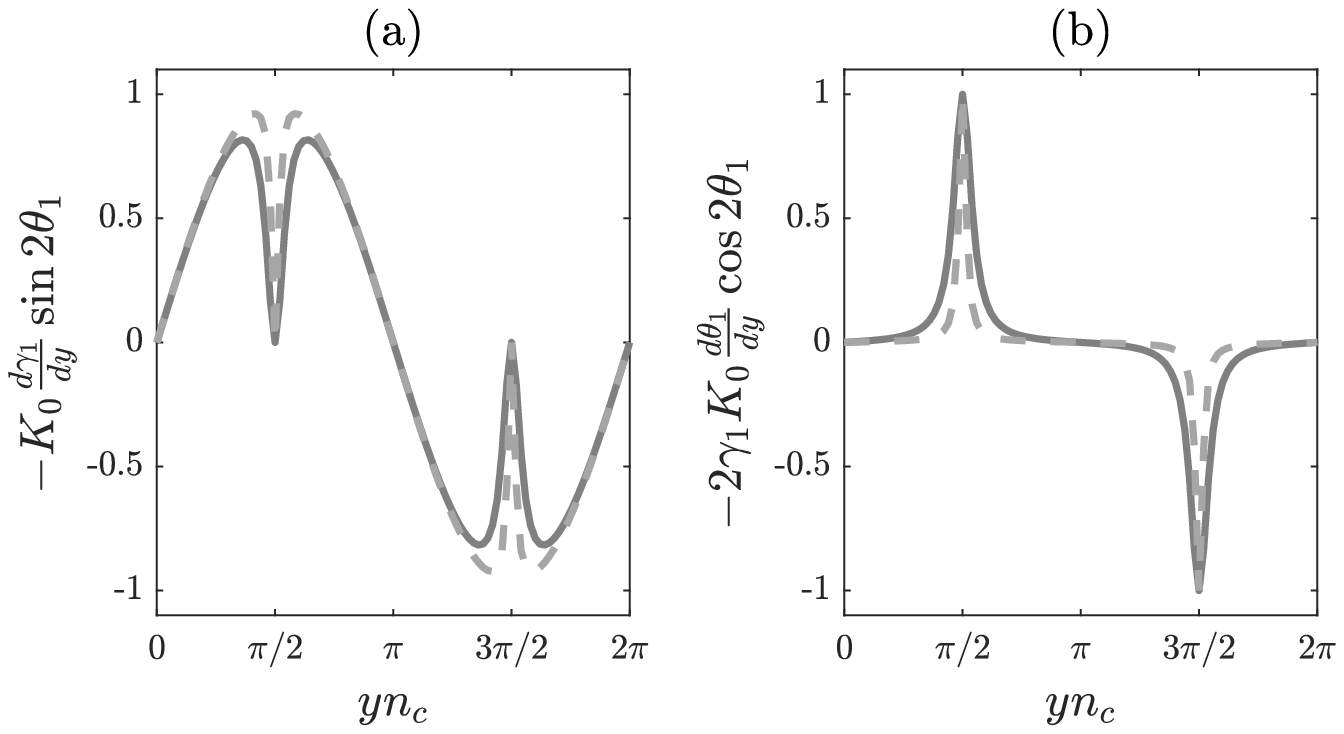}
\caption{{The contribution of (a) the anisotropy parameter gradient and (b) the tilt gradient to the acceleration of the most
unstable jet perturbation for $\beta=0.1$ (solid line) and $\beta=10$ (dashed line). The forcing is isotropic ($\eta=0$).}
\label{fig:jet_accel_contr}}
\end{figure}

The contribution of the gradients of each of the variance ellipse parameters in the jet acceleration is given to
leading order by (cf. Equation (\ref{eq:f1_iso})):
\begin{equation}
\overline{v'\zeta'}\simeq f_1=-K_0\frac{\partial\gamma_1}{\partial y}\sin 2\theta_1-2\gamma_1 K_0\frac{\partial\theta_1}{\partial y}\cos 2\theta_1,
\label{eq:vor_flux_iso}
\end{equation}
and is shown in Fig. \ref{fig:jet_accel_contr} for
$\beta=0.1$ and $\beta=10$. The first term of (\ref{eq:vor_flux2}) that involves the kinetic energy gradient is second order with respect to the
infinitesimal jet perturbation and is, therefore, negligible. We observe that the jet is intensified due to the gradient of the anisotropy parameter
$\gamma_1$ for most latitudes, while the gradient of the tilt contributes significantly only at the jet core.

\begin{figure}
\centerline{\includegraphics[width=35pc]{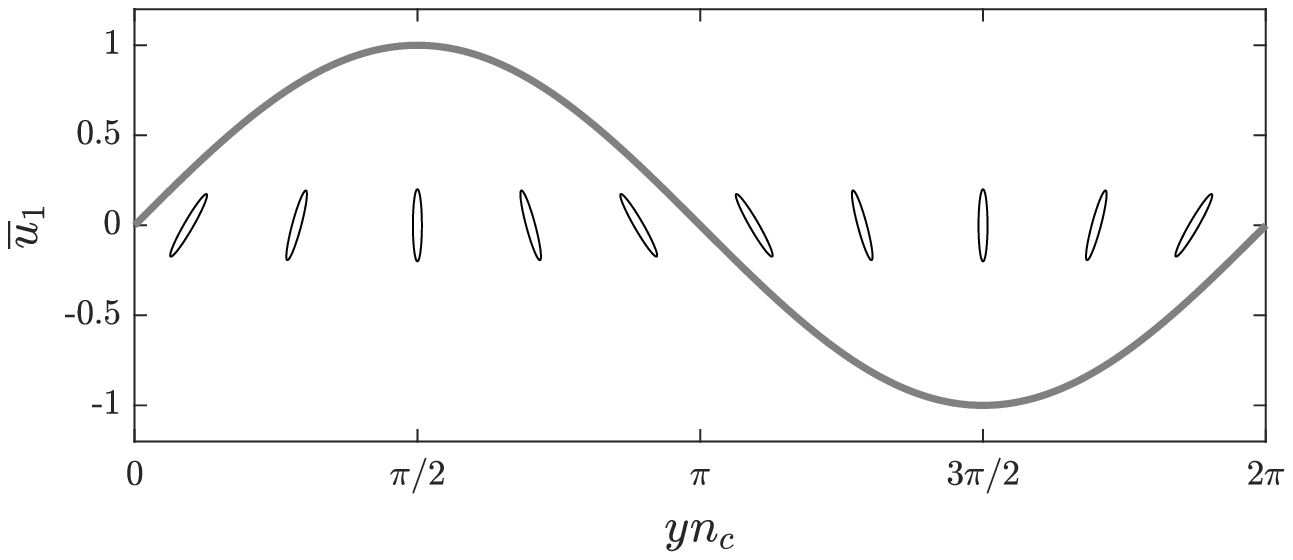}}
\caption{\label{fig:ellipse_noniso}Schematic change of the variance ellipse within the most unstable jet perturbation
shown by the thick line. The ellipse eccentricity is $1-\mathcal{O}(\mu)$, where $\mu\ll1$ is
the infinitesimal amplitude of the jet perturbation. So the ellipse eccentricity is set to $0.99$ for illustration purposes.
Similarly, changes in the tilt angle are of $\mathcal{O}(\mu)$, so changes in the angle are
exaggerated by $1/\mu$ for illustration purposes. The forcing is anisotropic ($\eta=1$).}
\end{figure}

For the anisotropic forcing, the eddy anisotropy and the eddy kinetic energy are to leading order equal to the values calculated for the homogeneous
equilibrium $M_0=K_0=\varepsilon_c/2$, while the eddy momentum fluxes are of the same order as the jet perturbation. The amplitude of the
fluxes that is shown in Fig. \ref{fig:fluxes}(c) is constant ($\hat{N}_1\simeq \sqrt{2}$) for low values of $\beta$, while for $\beta\gg1$ it
increases as $\hat{N}_1\simeq \left(3\beta/4\right)^{1/3}$. The eddy anisotropy parameter is to leading order:
\begin{equation}
\gamma_0=\frac{M_0}{K_0}=1,
\end{equation}
yielding a highly eccentric ellipse with eccentricity slightly smaller than one. The tilt angle
\begin{equation}
\theta_1=\frac{\pi}{2}-\frac{1}{2}\frac{\hat{N}_1\cos(n_cy)}{M_0},
\end{equation}
varies sinusoidally at leading order around $\pi/2$. This leads to the changes of the ellipse variance with latitude
that are schematically shown in Fig. \ref{fig:ellipse_noniso}. We observe again that the ellipse tilt is consistent
with the intuitive picture of the eddies leaning with the jet shear in order to produce upgradient fluxes. The ellipse
focal point and eccentricity determined by $M_0$ and $K_0$ are constant across latitude,
while the tilt increases in the first half of the jet wavelength. This corresponds to pattern E of Fig.
\ref{fig:ellipse-eastward-westward} leading to an eastward acceleration of the flow. In the second half of the jet wavelength, the
tilt decreases with latitude and corresponds to the opposite change than pattern E, leading to a westward acceleration of the jet.
Therefore the jet perturbation is reinforced. Since the terms of (\ref{eq:vor_flux2}) involving the kinetic energy and the anisotropy parameter
gradients are of second order with respect to the infinitesimal jet perturbation, the jet is intensified solely due to the gradient of the tilt
angle $\theta_1$. In this case the vorticity fluxes are to leading order given by (cf. Equation (\ref{eq:f1_noniso})):
\begin{equation}
\overline{v'\zeta'}\simeq f_1=2\gamma_0 K_0\frac{\partial\theta_1}{\partial y}=n_c\hat{N}_1\sin(n_cy).\label{eq:vflux_aniso}
\end{equation}
Since the jet is accelerated by the change of the ellipse tilt, we can parameterize this term. It was shown by
\cite{Bakas-Ioannou-2013-jas} that the eddy fluxes act as negative diffusion for $\beta\ll1$, while in the limit of $\beta\gg1$
the fluxes act as a constant term proportional to the jet velocity. Therefore, changes in the tilt in these limits can be parameterized
as proportional to the shear ($\theta\sim \pi/2-dU/dy$) and proportional to the integral of the jet velocity ($\theta\sim \pi/2+\int Udy$)
respectively.

\section{The eddy variance ellipse during the equilibration of the zonostrophic instability}

In this section we discuss the changes in the variance ellipse that bring about the equilibration of
zonostrophic instability. It can be readily shown that for energy input rates slightly above the critical
threshold $\epsilon=(1+\mu^2)\varepsilon_c$, where $\mu\ll1$, zonal jets with wavenumbers $|n-n_c|=\mathcal{O}(\mu)$ grow at a
rate $\s=\mathcal{O}(\mu^2)$. Guided by this observation, \cite{Parker-Krommes-2014-generation}  considered
that the jet velocity varies on the long time scale $T=\mu^2 t$ and is given by an expansion in $\mu$:
\begin{equation}
\overline{u}=\mu \overline{u}_1(y,T)+\mu^2\overline{u}(y,T)+\mu^3\overline{u}_3(y,T)+\mathcal{O}(\mu^4),\label{eq:uexp}
\end{equation}
with a similar expansion for the slowly varying eddy covariance
\begin{equation}
C= C_0({\bf x}_a-{\bf x}_b)+\mu C_1({\bf x}_a, {\bf x}_b,T)+\mu^2\,C_2({\bf x}_a, {\bf x}_b,T)+\mu^3\,C_3({\bf x}_a, {\bf x}_b,T)
+\mathcal{O}(\mu^4).\label{eq:Cexp}
\end{equation}
The order one terms yield the homogeneous equilibrium with $C_0=\varepsilon_cQ/2$, while the order $\mu$ terms
yield the eigenfunction (\ref{eq:eigfunc}) with amplitude $A(T)$:
\begin{eqnarray}
\overline{u}_1&=&A(T)e^{in_cy}+A^*(T)e^{-in_cy},\nonumber\\
C_1&=&\hat{C}_1^1e^{in_c(y_a+y_b)/2}+(\hat{C}_1^1)^*e^{-in_c(y_a+y_b)/2},
\label{eq:ordermu}
\end{eqnarray}
where $\hat{C}_1^1=A(T)\left[G^+({\bf x}_a-{\bf x}_b)-G^-({\bf x}_a-{\bf x}_b)\right]$. At order $\mu^2$, the quasi-linear
interaction of the infinitesimal jet $\um_1$ with the eddy covariance perturbation $C_1$ generates a
double-harmonic jet with velocity $\overline{u}_2=a_2A^2e^{2in_cy}+\mbox{c.c}$, where $a_2$ is an order one constant given
by (\ref{eq:a1}) and a corresponding covariance correction
\begin{equation}
C_2=\hat{C}_2^0+\hat{C}_2^2e^{2in_c(y_a+y_b)/2}+(\hat{C}_2^2)^*e^{-2in_c(y_a+y_b)/2}.\label{eq:C2}
\end{equation}
This contains a double-harmonic component to support the double-harmonic jet through the corresponding vorticity
fluxes and a homogeneous component that was shown by \cite{Bakas-etal-2019} to be a necessity of energy conservation: as the total
energy is conserved, the growing energy of the emerging jet has to be drained from the eddy energy and the homogeneous
correction facilitates this exchange.

At the next order, elimination of secular terms yields that the amplitude of the emerging jet satisfies the Ginzburg--Landau equation:
\begin{equation}
c_1\partial_TA=A-c_3|A|^2A,\label{eq:GLeq}
\end{equation}
where the coefficients $c_1$ and $c_3$ are given by (\ref{eq:c1}) and (\ref{eq:c3_ec}) and for the ring forcing considered here,
are real and positive. The linear term in (\ref{eq:GLeq}) drives the zonostrophic instability, while the non-linear term
is stabilizing and leads to its equilibration at the amplitude $A=1/\sqrt{c_3}$. It is shown in \cite{Bakas-etal-2019} that
the non-linear term comprises two parts $c_3=c_3^{ec}+c_3^{12}$. The first part is due to the interaction of the
first order jet $\overline{u}_1$ with the homogeneous correction to the covariance $\hat{C}_2^0$ and was referred to as
the energy correction term. The second part is due to the interaction of $\overline{u}_1$ with the double harmonic
$\hat{C}_2^2e^{2in_c(y_a+y_b)/2}$ and the interaction of $\overline{u}_2$ with $C_1$. We can therefore deduce which type
of eddy-mean flow interactions underlie the equilibration of the instability.

\begin{figure}
\centering\includegraphics[width=35pc]{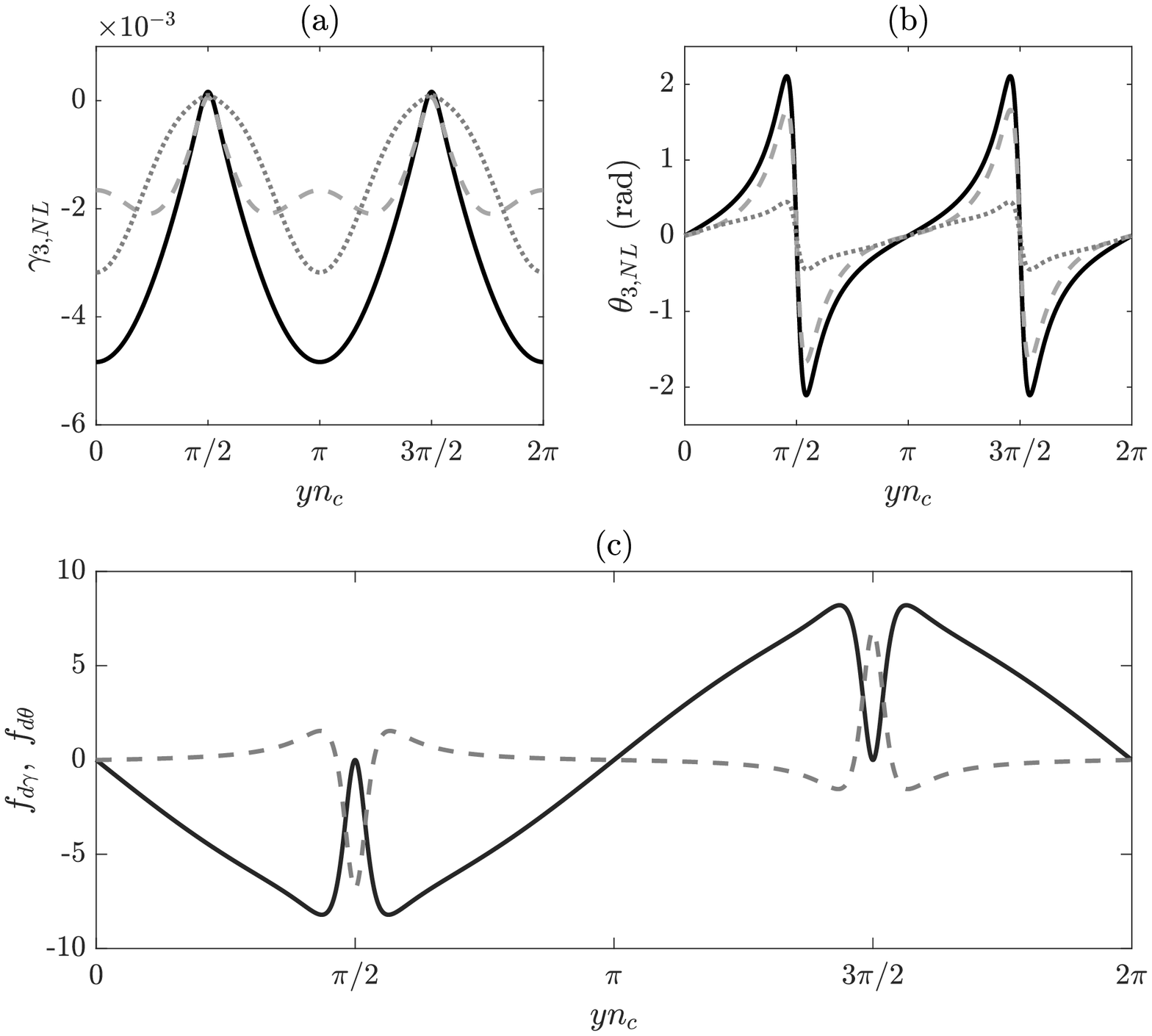}
\caption{The non-linear correction of (a) the tilt angle $\theta_{3,NL}$ and (b) the eddy anisotropy parameter $\gamma_{3,NL}$
as a function of latitude (solid lines). The contributions $\theta_{3,ec}$, $\gamma_{3,ec}$ (dashed line) and $\theta_{3,12}$,
$\gamma_{3,12}$ (dotted line) to the tilt and anisotropy parameter corrections are also shown. (c) The contribution of the
anisotropy parameter gradient $f_{d\gamma}$ (solid) and the tilt angle gradient
$f_{d\theta}$ (dashed) to the acceleration of the most unstable jet perturbation. The planetary vorticity gradient is $\beta=0.1$ and the forcing
is isotropic ($\eta=0$).}
\label{fig:NL_iso1}
\end{figure}

\begin{figure}
\centering\includegraphics[width=35pc]{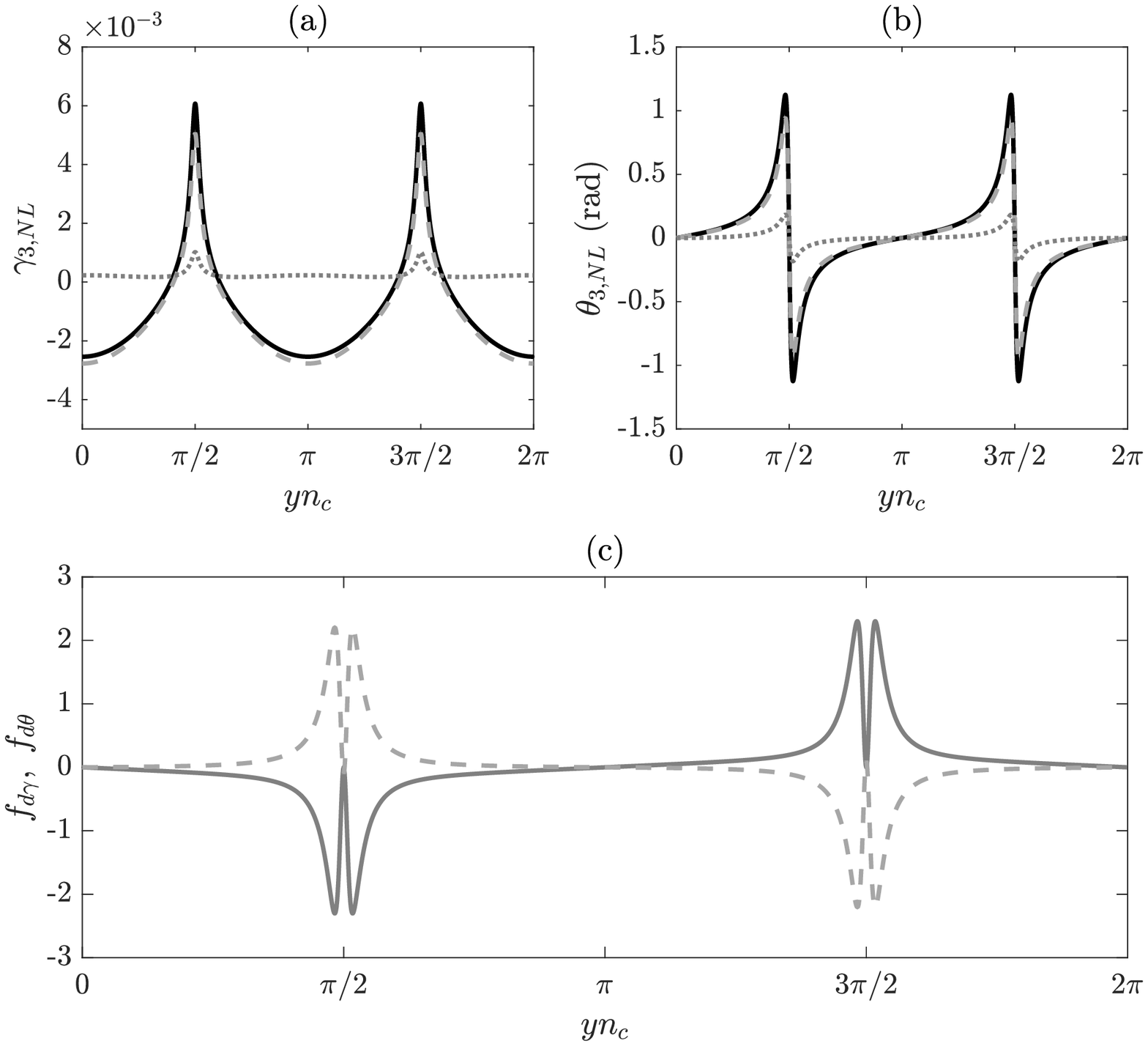}
\caption{The same as in Fig. \ref{fig:NL_iso1} but for $\beta=10$.}
\label{fig:NL_iso2}
\end{figure}

It is shown in Appendix B that for isotropic forcing,
the vorticity fluxes underlying the amplitude tendency in (\ref{eq:GLeq}) are (cf. Equation (\ref{eq:vorflux_final})):
\begin{eqnarray}
f_3&=&f_1\underbrace{-K_0\frac{\partial\gamma_{3,NL}}{\partial y}\sin 2\theta_1-2K_0\frac{\partial\gamma_1}{\partial y}\theta_{3,NL}\cos\ 2\theta_1}_{f_{d\gamma}}
\nonumber\\ &+&\underbrace{4\frac{\partial\theta_1}{\partial y}\gamma_1\theta_{3,NL}\sin 2\theta_1-2K_0\left(\frac{\partial\theta_{3,NL}}{\partial y}\gamma_1+\frac{\partial\theta_1}{\partial y}\gamma_{3,NL}\right)\cos\ 2\theta_1}_{f_{d\theta}}.\label{eq:vorflux_final_t}
\end{eqnarray}
The first term $f_1$, is the flux given by (\ref{eq:vor_flux_iso}) that underlies the linear term in (\ref{eq:GLeq}) and drives the
zonostrophic instability. The other two terms ($f_{d\gamma}$ and $f_{d\theta}$) underlie the non-linear stabilizing term with $\theta_{3,NL}$, $\gamma_{3,NL}$ given
by (\ref{eq:theta3NL}) and (\ref{eq:gamma3NL}) respectively. In the limit of $|A|\ll1$, these terms are $|A|^2$ smaller than the first term and
(\ref{eq:vorflux_final_t}) is to leading order
\begin{eqnarray}
f_3&\simeq& -K_0\left[\frac{\partial}{\partial y}\left(\gamma_1+\gamma_{3,NL}\right)\right]\sin \left(2\theta_1+2\theta_{3,NL}\right)\nonumber\\ &-&K_0(\gamma_1+\gamma_{3,NL})\left[\frac{\partial}{\partial y}\sin(2\theta_1+2\theta_{3,NL})\right].
\end{eqnarray}
Therefore in this limit, $\theta_{3,NL}$ and $\gamma_{3,NL}$ are the non-linear corrections to the tilt angle and the eddy anisotropy
parameter respectively. These corrections are shown in Fig. \ref{fig:NL_iso1} and Fig. \ref{fig:NL_iso2} for $\beta=0.1$ and
$\beta=10$ respectively. For the low value of $\beta$ (Fig. \ref{fig:NL_iso1}(a)-(b)), we observe that the anisotropy parameter is decreased
in the regions of maximum shear thereby decreasing the gradient of $\gamma$. The correction to the tilt angle has a dipole
structure reducing the gradient of the tilt at the jet core while increasing it in the regions of maximum shear. The contribution of the gradients of these corrections
($f_{d\gamma}$ and $f_{d\theta}$) to the vorticity fluxes are shown in Fig. \ref{fig:NL_iso1}(c). We can see that both
$f_{d\gamma}$ and $f_{d\theta}$ oppose the resulting contributions driving the instability and that the correction due to
the decrease of anisotropy is dominant except at the jet cores. Therefore in this limit, a partial isotropization of the eddies equilibrates
the instability. For the large value of $\beta$ (Fig. \ref{fig:NL_iso2}(a)-(b)), we observe that the gradient of the anisotropy parameter is
decreased through an increase of $\gamma$ at the jet cores and a decrease in the regions of maximum shear. The correction to the tilt angle
has the same dipole structure as in the low $\beta$ limit. The contribution of the gradients of these corrections
to the vorticity fluxes are shown in Fig. \ref{fig:NL_iso2}(c). We can see that both gradients are of the same order and
roughly follow the same pattern. As a result, the stabilizing fluxes are the small residual of these two opposing contributions.
The contribution of the interaction of the jet $\overline{u}_1$
with the homogeneous correction to the eddy covariance (termed as energy correction) and the interaction of the single harmonic
jet-eddy components with their double harmonic counterparts to the tilt angle ($\theta_{3,NL}$) and the anisotropy parameter
($\gamma_{3,NL}$) corrections are given by (\ref{eq:theta3NL}), (\ref{eq:gamma3NL}) and are shown in Figures
\ref{fig:NL_iso1}-\ref{fig:NL_iso2}. We observe that for low $\beta$ the energy correction interaction determines the angle correction while the interaction of the
single and the double harmonic jet-eddy components mostly determines the anisotropy parameter correction. Since
$f_{d\gamma}$ is the dominant contribution to the fluxes, the interaction of the
single and the double harmonic jet-eddy components underlies the stabilizing fluxes. This should be contrasted to the
corresponding contributions in the large $\beta$ limit shown in Fig. \ref{fig:NL_iso2}(a)-(b). We can see that in this limit the energy
correction interactions determine both the tilt and the anisotropy parameter corrections and therefore these interactions underlie the
stabilizing term.

For anisotropic forcing, the vorticity fluxes underlying the amplitude tendency in (\ref{eq:GLeq}) are:
\begin{equation}
f_3=f_1+2K_0\gamma_0\frac{\partial \theta_{3,NL}}{\partial y},\label{eq:vorflux_final_an_t}
\end{equation}
where $f_1$ is given by (\ref{eq:vflux_aniso}) and drives the instability and
\begin{equation}
\theta_{3,NL}=|A|^2\left(\frac{\pi}{2}-\theta_1\right),
\end{equation}
is the non-linear correction to the tilt angle. Therefore, the gradient of $\theta_{3,NL}$ opposes
the gradient of $\theta_1$ hindering the instability and at equilibrium it exactly cancels it out.
It is
also worth noting that since $\theta_{3,NL}$ is proportional to $\theta_1$, the parametrization of the angle discussed in the
previous section holds for the equilibration phase of the instability as well. The contribution of the energy correction interactions ($\theta_{3,ec}$) and the
interaction of the single harmonic jet-eddy components with their double harmonic counterparts to the tilt angle
($\theta_{3,12}$) are given by (\ref{eq:theta3NL_ani}). It is found that the energy correction interactions
dominate the contribution for all values of $\beta$ (not shown) and as a result these interactions underlie the stabilizing fluxes.

\section{Conclusion \label{sec:concl}}

We revisited the emergence of zonal jets in forced--dissipative barotropic beta-plane turbulence providing a direct link
between the eddy--mean flow dynamics underlying jet formation and the characteristics of the Reynolds stress tensor. This
tensor is the central object of the GEOMETRIC approach, a framework for studying and parameterizing the effect of small--scale
eddies on the large--scale flow. To address this, we employed the statistical state dynamics of the turbulent flow closed at
second order as it provides both an analytic expression for the zonostrophic instability (that was shown by previous studies to
form the zonal jets) and an analytic expression for the stress tensor.

Zonostrophic instability arises as the homogeneous statistical equilibrium turbulent state becomes unstable when the energy input
rate of the forcing passes a critical value. In order to address the linear phase of the instability, the stress tensor was
calculated at the stability boundary, that is when the energy input rate assumes this critical value both for isotropic forcing and
for anisotropic forcing mimicking baroclinic instability. The stress tensor, which
is visualized by an eddy variance ellipse, was rewritten in terms of two bounded parameters, the ellipse tilt and
the anisotropy parameter determining the ellipse eccentricity, as well as the eddy kinetic energy determining the size
of the ellipse. The change of the ellipse characteristics were then linked to the eddy vorticity fluxes that form the jets.

We
found that for isotropic forcing, changes in the eddy kinetic energy are of higher order and
do not influence the emerging jet. Regarding the tilt of the ellipse, we found that the eddies tilt
with the shear surrendering their energy to the mean flow as would be intuitively expected for jet formation  with
the tilt angle being piecewise constant. Therefore, the gradient of the
tilt that contributes directly to jet acceleration is very small except at the jet cores. In the bulk of the flow, the jet is
accelerated as the infinitesimal jet perturbation anisotropizes the eddies in such a way as to produce upgradient fluxes. The
anisotropy parameter is maximum at regions of large shear and minimum at the jet core. For anisotropic
forcing, both the eddy kinetic energy and the anisotropy parameter are of higher order and the jet is accelerated solely due to changes
in the ellipse tilt. The change in the tilt can be parameterized in this case as proportional to the shear for low values of the
planetary vorticity gradient or as the integral of the flow for large values of the planetary vorticity gradient, a parametrization
that was shown to hold even in the equilibration stage of the instability.

The non--linear phase of zonostrophic instability and its equilibration were then studied in the limit in which the energy input rate
is slightly above the critical value for jet formation. In this limit, the amplitude of the jet velocity follows the weakly non--linear
Ginzburg--Landau dynamics therefore allowing us to obtain analytic expressions for the stress tensor and its characteristics and to
identify the two types of interaction that underlie the equilibration. The first is the interaction of the infinitesimal jet with the
homogeneous correction to the eddy covariance that ensures total energy conservation and was termed as energy correction by \cite{Bakas-etal-2019}.
The second is the interaction of the single harmonic
jet-eddy components with their double harmonic counterparts that are generated at higher order. For isotropic forcing, the anisotropy parameter
is decreased in the regions of maximum shear during the non--linear phase thereby
isotropizing the eddies. For low values of the planetary vorticity gradient this is achieved
by the interaction of the single harmonic jet-eddy components with their double harmonic counterparts, while for large
values of the planetary vorticity gradient this is achieved by the energy correction interactions. The energy correction interactions
decrease the gradient of the tilt angle at the jet core and increase it in the regions of maximum shear. For low values of the planetary
vorticity gradient, the instability equilibrates due to
the partial isotropization of the eddies, while for large values of the planetary vorticity gradient the jet equilibrates at its
finite amplitude due to opposing gradients from both of the tilt and the anisotropy parameter. For anisotropic
forcing, the energy correction interactions simply reduce the tilt changes of the linear phase of the instability and
equilibrate the flow. In summary,
the geometric decomposition of the stress tensor in zonostrophic instability showed that either the anisotropization of the
eddies (for isotropic forcing) or the change of the eddy tilt (for anisotropic forcing) drive the emerging jet and that the
instability equilibrates as these changes are partially reversed by the non--linear jet-eddy dynamics.

\acknowledgments
The authors would like to thank Navid Constantinou for helpful comments on the manuscript.


\appendix[A]

\appendixtitle{Calculation of the stress tensor at the onset of zonostrophic instability}
\label{app:dispersion}
In this Appendix we calculate the stress tensor at the onset of zonostrophic instability. We first calculate
the most unstable mode with wavenumber $n_c$ yielding the maximum vorticity fluxes and the critical energy
input rate $\varepsilon_c$ above which this mode becomes unstable and jets form. The vorticity fluxes
induced by the eigenfunction (\ref{eq:eigfunc}) are \citep{Srinivasan-Young-2012,Bakas-etal-2015}:
\begin{eqnarray}
& &\overline{v'\zeta'}=\Rcal\left(\delta C\right)=\epsilon f(\sigma|\delta \overline{u}, Q/2)e^{iny}
\nonumber\\&=&\epsilon e^{iny}\int\frac{d^2{\bf k}}{2\pi}\frac{nk_x^2(k_y+ n /2)\,(1- n ^2/k^2)}{(\sigma+2)
k^2k_{s}^2+2i\beta  n  k_x (k_y+ n /2)}\hat{Q}({\bf k})\ ,\label{eq:ff}
\end{eqnarray}
with $k_s=|{\bf k}+{\bf n}|$. In order to calculate $n_c$, $\epsilon_c$, we
assume criticality ($\sigma=0$), substitute the ring forcing power spectrum (\ref{eq:Qf}),
express the integrand in polar coordinates $(k_x,k_y)=(k\cos\phi, k\sin\phi)$ and integrate
over $k$ to obtain:
\begin{equation}
\hat{f}_1=\int_0^{2\pi}\frac{N_f}{D_f+i\beta D_\beta}d\phi\ ,\label{eq:vorflux}
\end{equation}
with $N_f= n \cos^2\phi(\sin\phi+ n /2)(1- n ^2)\left[1+\eta\cos(2\phi)\right]/ 2\pi$,
$D_f=1+2n\sin\phi+ n^2$ and $D_\beta=n \cos\phi(\sin\phi+ n/2)$. By numerically
calculating the integral and maximizing its value over $n$, we identify the most unstable
jet with wavenumber $n_c$. The energy input rate $\varepsilon_c$ is then given by (\ref{eq:stab_bound}).
The integral is real with $\hat{f}_1(-n)=\hat{f}(n)$. Therefore, for the
sinusoidal jet $\overline{u}_1=\sin(n_cy)$ the vorticity fluxes at the onset are:
\begin{equation}
\overline{v'\zeta'}=\mbox{Im}\left(\varepsilon_c \hat{f}_1(n_c)e^{in_cy}\right)=\varepsilon_c\hat{f}_1(n_c)\sin(n_cy).
\end{equation}

We then calculate the elements $N$ and $M$ of the stress tensor as well as the eddy kinetic energy $K$. The momentum fluxes $N$
are given by the action of the operator $\mathcal{N}$ on the covariance:
\begin{eqnarray}
N&=&\mathcal{N}\left(C_0+\delta C\right)\nonumber\\ &\defn& -\frac{1}{2}\[\(\partial_{x_a}\partial_{y_b}+\partial_{x_b}\partial_{y_a}\)
\Del_a^{-1}\Del_b^{-1}\left(C_0+\delta C\right)\]_{a=b}.
\end{eqnarray}
After substitution of the eigenfunction (\ref{eq:eigfunc}), we find that the induced fluxes for the sinusoidal jet
perturbation $\overline{u}_1$ are:
\begin{equation}
N=\mbox{Im}\left[\frac{i\epsilon}{n}f(\sigma|\delta \overline{u}, Q/2)e^{iny}\right]=\hat{N}_1\cos(ny),\label{eq:mf}
\end{equation}
where $\hat{N}_1=\hat{f}_1/n$.

Similarly, $M$ and $K$ are given
by the action of the operators $\mathcal{M}$ and $\mathcal{K}$ on the covariance:
\begin{eqnarray}
M&=&\mathcal{M}\left(C_0+\delta C\right)\nonumber\\ &\defn&\frac{1}{4}\[\(\partial_{x_a}\partial_{x_b}-\partial_{y_a}\partial_{y_b}\)\Del_a^{-1}\Del_b^{-1}
\left(C_0+\delta C\right)\]_{a=b},
\end{eqnarray}
\begin{eqnarray}
K&=&\mathcal{K}\left(C_0+\delta C\right)\nonumber\\ &\defn&\frac{1}{4}\[\(\partial_{x_a}\partial_{x_b}+\partial_{y_a}\partial_{y_b}\)\Del_a^{-1}\Del_b^{-1}
\left(C_0+\delta C\right)\]_{a=b}.
\end{eqnarray}
For the jet perturbation $\overline{u}_1$ they are:
\begin{eqnarray}
M&=&M_0+\hat{M}_1\sin(ny)=\frac{\epsilon}{8\pi}\int d^2{\bf k}(k_x^2-k_y^2)\hat{Q}({\bf k})\nonumber\\&+&
\epsilon \sin(ny)\int\frac{d^2{\bf k}}{4\pi}\frac{ik_x(k_x^2-k_y^2-nk_y)(1- n ^2/k^2)}{(\sigma+2)
k^2k_{s}^2+2i\beta  n  k_x (k_y+ n /2)}\hat{Q}({\bf k}),\label{eq:eddyan}
\end{eqnarray}
\begin{eqnarray}
K&=&K_0+\hat{K}_1\sin(iny)=\frac{\epsilon}{8\pi}\int d^2{\bf k}(k_x^2+k_y^2)\hat{Q}({\bf k})\nonumber\\ &+&
\epsilon \sin(iny)\int\frac{d^2{\bf k}}{4\pi}\frac{ik_x(k_x^2+k_y^2-nk_y)(1- n ^2/k^2)}{(\sigma+2)
k^2k_{s}^2+2i\beta  n  k_x (k_y+ n /2)}\hat{Q}({\bf k}).\label{eq:EKE}
\end{eqnarray}
To obtain the integrals we substitute $\hat{Q}$, move to polar coordinates as above and integrate
over $k$ and $\phi$. The first integrals yield $M_0=\varepsilon\eta/2$ and $K_0=\varepsilon/2$. The second
term in (\ref{eq:EKE}) is a small correction that does not affect the jet acceleration and is therefore not calculated,
while the second term in (\ref{eq:eddyan}) at the onset of
instability is:
\begin{equation}
\hat{M}_1=\int_0^{2\pi}\frac{iN_M}{D_f+i\beta D_\beta}d\phi\ ,\label{eq:Mc}
\end{equation}
with $N_M= \epsilon\cos\phi(\cos 2\phi-n\sin\phi)(1- n ^2)\left[1+\eta\cos(2\phi)\right]/ 4\pi$.
Therefore by numerically evaluating the integrals (\ref{eq:vorflux}) and (\ref{eq:Mc}) for the most
unstable jet with wavenumber $n=n_c$ and for the critical rate $\varepsilon=\varepsilon_c$ we calculate the
stress tensor at the onset of instability.

We then estimate asymptotic values for the integrals and the stress tensor in the limits of both
small and large values of the planetary vorticity gradient. In the limit $\beta\ll1$ we expand the integrand in
(\ref{eq:vorflux}) in powers of $\beta$ to obtain:
\begin{equation}
\hat{f}_1=\sum_{j=1}^\infty (-i\beta)^j\int_0^{2\pi}\frac{N_fD_\beta^{j-1}}{D_f^j}d\phi.
\end{equation}
The integrals of the odd powers of $\beta$ are zero, while the rest can be evaluated analytically through
contour integration. For anisotropic forcing ($\eta=1$) we obtain $\hat{f}_1=n^2(1-n^2)/8+\mathcal{O}(\beta^2)$. The maximum
value of $\hat{f}_1$ over the jet wavenumber $n$ is attained at $n_c=1/\sqrt{2}$. The critical energy input
rate is then according to (\ref{eq:stab_bound}) $\epsilon_c=1/\hat{f}_1=32$. For isotropic forcing ($\eta=0$) we have:
\begin{equation}
\hat{f}_1=\frac{\beta^2n^4}{1024}\left(48+\frac{5\beta^2n^2(n^4+2n^2-4)}{(1-n^2)^2}\right)+\mathcal{O}(\beta^6).
\end{equation}
The maximum value of $\hat{f}_1$ can be evaluated approximately to occur for $n_c=1-\left(10\beta^2/768\right)^{1/3}+\mathcal{O}(\beta^4)$,
yielding a critical energy input rate $\epsilon_c\simeq 64/3\beta^2$. Therefore, from (\ref{eq:mf}) we
calculate the momentum fluxes at the onset:
\begin{equation}
\hat{N}_1\simeq 1+(\sqrt{2}-1)\eta.\label{eq:Nbs}
\end{equation}
Similarly for the eddy anisotropy, we expand the integrand in (\ref{eq:eddyan}) in powers of $\beta$. The non-zero leading
order integral for $\eta=0$ is:
\begin{equation}
\hat{M}_1\simeq\frac{\epsilon\beta^2n^3(1-n^2)}{32\pi}\int_0^{2\pi}\frac{\cos^4\phi(\cos 2\phi-n\sin\phi)
(8\sin^3\phi+12n\sin^2\phi+6n^2\sin\phi+n^3)}{(1+2n\sin\phi+n^2)^4}d\phi,
\end{equation}
which upon contour integration yields $\hat{M}_1=\epsilon\beta^3n^6/128(1-n^2)$. At the onset of instability ($\epsilon=\epsilon_c$)
and for the most unstable jet ($n=n_c$) the eddy anisotropy is:
\begin{equation}
\hat{M}_1=\left(\frac{12\beta}{270}\right)^{1/3}.\label{eq:Mbs}
\end{equation}

In the limit $\beta\gg1$, the most unstable jet has small wavenumber \citep{Srinivasan-Young-2012}. We can therefore obtain
an asymptotic expression for $n_c$ and $\epsilon_c$ by expanding the
integrand in (\ref{eq:vorflux}) in powers of $n$ but keeping $\beta n\sim\mathcal{O}(1)$ (since $\beta\gg1$) to obtain:
\begin{eqnarray}
\hat{f}_1&=&\frac{n^2(1-n^2)}{4\pi}\int_0^{2\pi}\frac{\cos^2\phi(1-4\sin^2\phi)(1+\eta\cos 2\phi)}{(1+i\beta n\cos\phi\sin\phi)^2}d\phi
\nonumber\\ &+&
\mathcal{O}(n^4).
\end{eqnarray}
The integral can be evaluated analytically through contour integration yielding:
\begin{equation}
\hat{f}_1=\frac{(1-n^2)\left\{-(16+6\beta^2n^2)+(8+2\beta^2n^2)\left[(1+\eta/2)\sqrt{4+\beta^2n^2}-\eta\right]\right\}}{\beta^2(4+\beta^2n^2)^{3/2}}.
\end{equation}
The maximum value of $\hat{f}_1$ over the jet wavenumber $n$ is approximately attained at
$n_c\simeq \left[(3+\eta)/(2+\eta)\beta\right]^{1/3}$ and the critical energy input rate is
$\epsilon_c\simeq\beta^2/(2+\eta)$. The momentum fluxes
are calculated from (\ref{eq:mf}):
\begin{equation}
\hat{N}_1=\left(\frac{(2+\eta)\beta}{3+\eta}\right)^{1/3}.\label{eq:Nbl}
\end{equation}
To obtain $\hat{M}_1$ which is real, we rewrite (\ref{eq:Mc}) in the form:
\begin{equation}
\hat{M}_1=\frac{1}{\beta}\int_0^{2\pi}\frac{iN_M}{\chi D_f+iD_\beta}d\phi\ ,\label{eq:Mc1}
\end{equation}
where $\chi=1/\beta$ and expand the integrand in powers of $\chi$. The leading order real term is:
\begin{equation}
\hat{M}_1\simeq\frac{1}{\beta}\int_0^{2\pi}\frac{N_MD_\beta}{\chi^2 D_f^2+D_\beta^2}d\phi=\frac{1}{\beta}\int_0^{2\pi}F_\chi(\phi,n)d\phi\ .\label{eq:Mc2}
\end{equation}
For the angles $\phi$ for which $D_\beta\sim\mathcal{O}(1)$, the integrand is order one. However, if
$D_\beta\sim\mathcal{O}(\beta^{-1})$ for some angle $\phi_r$, then the denominator is order $\chi$ and
the integrand is large. The angles $\phi_r$ are the roots of $D_\beta$ which are at
$\phi_1=\pi/2$, $\phi_3=3\pi/2$ and the two angles satisfying $\phi_{2,4}=-\sin^{-1}(n/2)$.

To calculate asymptotic approximations to the integral $\hat{M}_1$, we split the range of integration to a small range
close to the roots of $D_\beta$ for which the integrand is large, $I^{(\textrm{R})}$, and to a range away from the roots
of $D_\beta$, $I^{\textrm{(NR)}}$:
\begin{equation}
\hat{M}_1 = \sum_{j=1}^4\[ \spb \right. \underbrace{\int\limits_{\phi_{j-1}+\delta\phi}^{\phi_j-\delta\phi} \!\!
F_\chi(\phi,n)\,d\phi}_{I^{(\textrm{NR})}_j}  + \underbrace{\int
\limits_{\phi_{j}-\delta\phi}^{\phi_j+\delta\phi} \!\! F_\chi(\phi,n)\,d\phi}_{I^{(\textrm{R})}_j}
\left.\spb \]~,\label{eq:sumI}
\end{equation}
where the range $\delta\phi=a\chi$, with $a$ an order one parameter. Asymptotic approximations to
the integral over the two ranges are then found separately (cf.~\citet{Hinch-1991}).

To calculate $I^{(\textrm{R})}$, we rescale the angle close to the roots $\phi=\phi_j+a\chi u$, to obtain:
\begin{equation}
I^{(\textrm{R})}_j=\chi\int\limits_{-a}^{a}\frac{N_M(\phi_j+a\chi u) D_\beta(\phi_j+a\chi u)}
{D_f^2(\phi_j+a\chi u)+\chi^2D_\beta^2(\phi_j+a\chi u)}du+\mathcal{O}(\chi^{-3})~.
\end{equation}
We then Taylor expand the integrand in powers of $\chi$:
\begin{eqnarray}
& &I^{(\textrm{R})}_j=\large\int\limits_{-a}^{a}\frac{N_{M,j}D_{\beta,j}'u}
{D_{f,j}^2+{D_{\beta,j}'}^2u^2}du\nonumber\\
&+&\chi
\int\limits_{-a}^{a}\frac{\left[D_{f,j}^2\left(2D_{\beta,j}'N_{M,j}'+N_{M,j}D_{\beta,j}''\right)-
4D_{f,j}D_{f,j}'D_{\beta,j}'N_{M,j}\right]u^2}
{2\left(D_{f,j}^2+{D_{\beta,j}'}^2u^2\right)^2}du\nonumber\\ &+&\chi
\int\limits_{-a}^{a}\frac{{D_{\beta,j}'}^2\left(2D_{\beta,j}'N_{M,j}'-D_{\beta,j}''N_{M,j}\right)u^4}
{2\left(D_{f,j}^2+{D_{\beta,j}'}^2u^2\right)^2}du+\mathcal{O}(\chi^2),
\end{eqnarray}
where primed quantities are derivatives with respect to angle and the subscript $j$ denotes the value of the relative
function at $\phi_j$. Evaluation of the integrals and summation over all roots yields:

\begin{equation}
I^{(\textrm{R})} = \frac{\epsilon a\chi(1-n^2)^2}{4\pi(1-n^2/4)}+\frac{\epsilon\chi |n|(1-n^2)(n^2-2)}{8(1-n^2/4)^2}~.\label{eq:IR}
\end{equation}

\begin{figure}
\centerline{\includegraphics[width=20pc]{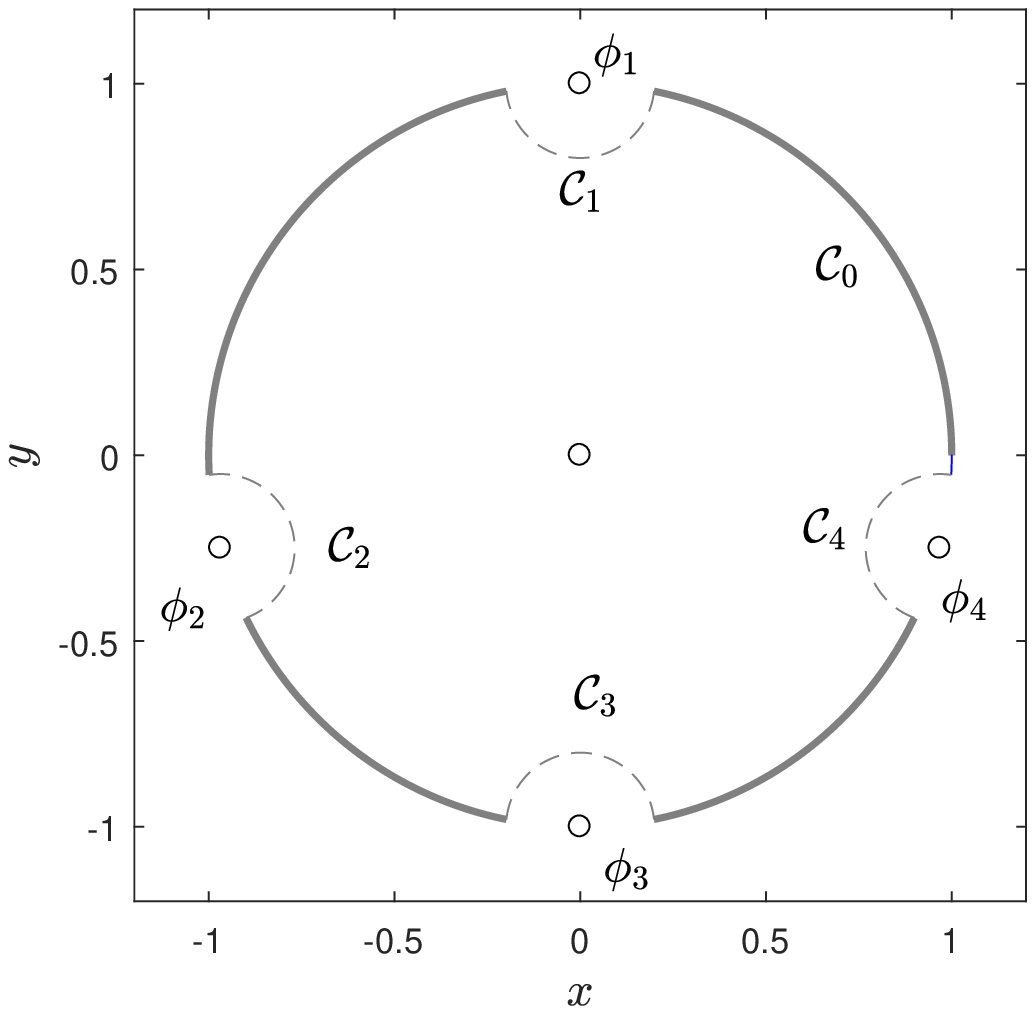}}
\caption{\label{fig:contour_int} Path of integration to calculate $I^{(\mathrm{NR})}$. Shown
is the path $\mathcal{C}_0$ (solid line) over which the integral is calculated and the complimentary paths $\mathcal{C}_j$ (dashed lines)
with $j=1,...,4$ around the angles $\phi_j$ (circles) with the help of which the integral is calculated. The pole of the integrand
at $z=0$ is also shown (circle).}
\end{figure}

To calculate $I^{(\textrm{NR})}$, we expand the integrand for $\chi\ll1$, to obtain to leading order:
\begin{equation}
I^{(\textrm{NR})}=\sum_{j=1}^4\int\limits_{\phi_{j-1}+\delta\phi}^{\phi_j-\delta\phi} \frac{N_M}{D_\beta}d\phi+\mathcal{O}(\chi^2)~.\label{eq:dvzR_largeb}
\end{equation}
To evaluate the sum of the integrals, we move into the complex plane by setting $z=e^{i\phi}$. The path of integration
$\mathcal{C}_0$ that lies on the unit circle is shown by the thick line in Fig. \ref{fig:contour_int} and excludes the region of
width $2\delta\phi$ close to the angles $\phi_j$. To evaluate the integrals,
we add the small paths $\mathcal{C}_j$, with $j=1,...,4$ around the angles $\phi_j$ that are also shown in
Fig. \ref{fig:contour_int}, to obtain a closed path
$\Sigma\mathcal{C}$. The paths $\mathcal{C}_j$ are defined by $z=e^{i\phi_j}+Re^{i\thet}$, where
$R=\sin\delta\phi/\sin(\pi+\omega)$, and
$\omega=\cot^{-1}\left[(\cos\delta\phi-1)/\sin\delta\phi\right]$. The angle $\thet$ is
limited by $|\thet-\phi_j-\pi|\leq\omega$.  The integrand has poles at the
angles $\phi_j$ that are outside the closed path and at $z=0$, so
by Cauchy's theorem the integral over the closed path $\Sigma\mathcal{C}$ is:
\begin{equation}
\int_{\Sigma\mathcal{C}}\frac{N_M}{izD_\beta}dz=2\pi\mathrm{Res}(0).
\end{equation}
The residue is zero yielding:
\begin{equation}
I^{(\textrm{NR})}=\int_{\mathcal{C}_0}\frac{N_M}{izD_\beta}dz=-\sum_{j=1}^4\int_{\mathcal{C}_j}\frac{N_M}{izD_\beta}dz.
\end{equation}
Evaluating the integrands at the paths $\mathcal{C}_j$, expanding the functions in powers of $\delta\phi$ and
calculating the resulting integrals yields:
\begin{equation}
I^{(\textrm{NR})} = -\frac{\epsilon\delta\phi(1-n^2)^2}{4\pi(1-n^2/4)}~.\label{eq:INR}
\end{equation}
Adding (\ref{eq:IR}) and (\ref{eq:INR}) yields $\hat{M}_1=\epsilon |n|(1-n^2)(n^2-2)/8\beta^2(1-n^2/4)^2$. At
the onset of zonostrophic instability $n=n_c$ and:
\begin{equation}
\hat{M}_1\simeq \left(\frac{3}{2^{10}\beta}\right)^{1/3}.\label{eq:Mbl}
\end{equation}

\appendix[B]

\appendixtitle{Calculation of the eddy variance ellipse during the equilibration of zonostrophic instability}

In this Appendix we study the equilibration of zonostrophic instability and calculate how
the variance ellipse changes during the development of the instability. To obtain the weakly non-linear dynamics governing the jet evolution near the onset of zonostrophic
instability, we assume that the energy input rate is slightly supercritical
$\e=\e_c(1+\mu^2)$, where $\mu\ll 1$. Following the discussion in section 5 we assume that
the most unstable jet with wavenumber $n_c$ evolves on a slow time scale $T=\mu^2t$:
$\um_1=A(T)e^{in_cy}+A^*(T)e^{-in_cy}$. We expand the velocity and the covariance as a series in
$\mu$ (Eqs. (\ref{eq:uexp})-(\ref{eq:Cexp})), substitute (\ref{eq:uexp})-(\ref{eq:Cexp}) in
(\ref{eq:cum1})-(\ref{eq:cum2}) and collect terms with equal powers of $\mu$. Details on the
calculation can be found in \cite{Bakas-etal-2019}. Here, we will only
recite the main results.

As described in section 5, the order one terms yield the homogeneous equilibrium with
$C_0=\varepsilon_cQ/2$, while the order $\mu$ terms yield the eigenfunction (\ref{eq:ordermu}). At order $\mu^2$, the
quasi-linear interaction of the infinitesimal jet $\um_1$
with the perturbation in the eddy covariance $C_1$ generates a covariance correction with a homogeneous component and a component at
the double harmonic $2n_c$ that is given by (\ref{eq:C2}). The homogeneous component
$\hat{C}_2^0({\bf x}_a-{\bf x}_b,T)=\varepsilon_cQ/2+\hat{C}_2^{0,NL}({\bf x}_a-{\bf x}_b,T)$
contains two parts. The first part is the contribution of the homogeneous
covariance $C_0=(1+\mu^2)\varepsilon_cQ/2$ at order $\mu^2$. The second part
\begin{eqnarray}
& &\hat{C}_2^{0,NL}({\bf x}_a-{\bf x}_b,T)=\mathcal{L}^{-1}\mathcal{J}_{Ae^{in_cy}}(\hat{C}_1^1)^*e^{-in_c(y_a+y_b)/2}\nonumber\\
&+&\mathcal{L}^{-1}\mathcal{J}_{A^*e^{-in_cy}}\hat{C}_1^1e^{in_c(y_a+y_b)/2},\label{eq:C2inhom}
\end{eqnarray}
is due to the quasi-linear interaction of $\overline{u}_1$ with $C_1$. In (\ref{eq:C2inhom}),
the operators $\mathcal{L}=-\beta\left(\partial_{x_a}\Delta_a^{-1}+\partial_{x_b}\Delta_b^{-1}\right)-2$ and $\mathcal{J}_{\overline{u}}=-\overline{u}_a\partial_{x_a}-\overline{u}_b\partial_{x_b}+\partial_{y_ay_a}^2\overline{u}_a\partial_{x_a}\Delta_a^{-1}+
\partial_{y_by_b}^2\overline{u}_b\partial_{x_b}\Delta_b^{-1}$ are the linear and non-linear parts of the operator $\mathcal{A}_a+\mathcal{A}_b$.
The double harmonic component drives a jet with velocity $\overline{u}_2=\alpha_2A^2(T)e^{2incy}+\mbox{c.c.}$ and is given by
\begin{eqnarray}
\hat{C}_2^2&=&e^{-2in_c(y_a+y_b)/2}\mathcal{L}^{-1}\mathcal{J}_{Ae^{in_cy}}\hat{C}_1^1e^{in_c(y_a+y_b)/2}
\nonumber\\ &+&e^{-2in_c(y_a+y_b)/2}\mathcal{L}^{-1}\mathcal{J}_{\alpha_2A^2e^{2in_cy}}(\varepsilon_cQ/2).
\end{eqnarray}
The constant $\alpha_2$ is an order one parameter given by:
\begin{equation}
\alpha_2=\dfrac{\dfrac{\varepsilon_c}{2} \displaystyle\int\dfrac{d^2{\bf k}}{2\pi}\dfrac{in_ck_x^3(k^2-n_c^2)}{k^2k^2_2+i\beta n_ck_xk_{y,1}} \left\{\dfrac{k_{y,2} (k_2^2- n_c ^2)}{k^2k_4^2+2i\beta n_c k_x k_{y,2} }-\dfrac{ k_y k^2_2 (k^2- n_c ^2)} {k^2( k_{-2}^2 k_2^2+2i\beta n_c k_x k_y)}\right\}\hat{Q}({\bf k})}{\varepsilon_c\displaystyle\int\dfrac{d^2{\bf k}}{2\pi}\dfrac{n_ck_x^2k_{y,2}(k^2-4 n_c ^2)}{k^2( k^2 k_4^2+2i\beta n_c k_x k_{y,2})}\hat{Q}({\bf k})-1}\ ,\label{eq:a1}
\end{equation}
where $k_{y,j}\equiv k_y + j n_c /2$ and $k_j^2\equiv k_x^2+k_{y,j}^2$ for any integer $j$ (eg. Equation (B7) of
\cite{Bakas-etal-2019}).

At order $\mu^3$, the quasi-linear interaction between the double harmonic jet $\overline{u}_2$ and the eddy covariance
$C_1$ as well as the interaction between $\overline{u}_1$ and $C_2$ generate the covariance correction
\begin{equation}
C_3=\hat{C}_3^1e^{in_c(y_a+y_b)/2}+\hat{C}_3^3e^{3in_c(y_a+y_b)/2}+\mbox{c.c.}.
\end{equation}
The $3n_c$ component drives a jet with velocity $\overline{u}_3=\alpha_3|A|^2Ae^{3in_cy}+\mbox{c.c}$ and is given by
\begin{eqnarray}
\hat{C}_3^3&=&e^{-3in_c(y_a+y_b)/2}\mathcal{L}^{-1}\mathcal{J}_{Ae^{in_cy}}\hat{C}_2^2e^{2in_c(y_a+y_b)/2}\nonumber\\
&+&e^{-3in_c(y_a+y_b)/2}\mathcal{L}^{-1}\mathcal{J}_{\alpha_3|A|^2Ae^{3in_cy}}(\varepsilon_cQ/2).
\end{eqnarray}
The constant $\alpha_3$ is an order one parameter. The component
\begin{equation}
\hat{C}_3^1({\bf x}_a-{\bf x}_b,T)=\hat{C}_1^{1,h}({\bf x}_a-{\bf x}_b,T)+\underbrace{\hat{C}_3^{1,ec}({\bf x}_a-{\bf x}_b,T)+
\hat{C}_3^{1,12}({\bf x}_a-{\bf x}_b,T)}_{\hat{C}_3^{1,NL}},
\end{equation}
contains three parts. The first part
\begin{equation}
\hat{C}_3^{1,h}=e^{-in_c(y_a+y_b)/2}\mathcal{L}^{-1}\mathcal{J}_{Ae^{in_cy}}(\varepsilon_cQ/2),\label{eq:C3hom}
\end{equation}
is due to the interaction of $\overline{u}_1$ with the homogeneous covariance. The second part is
\begin{equation}
\hat{C}_3^{1,ec}=e^{-in_c(y_a+y_b)/2}\mathcal{L}^{-1}\mathcal{J}_{Ae^{in_cy}}\hat{C}_0^{2,NL},
\end{equation}
is due to the quasi-linear interactions of $\overline{u}_1$ with the homogeneous correction $\hat{C}_2^0$ and is
referred in \cite{Bakas-etal-2019} as the energy correction term. The third part
\begin{eqnarray}
\hat{C}_3^{1,12}&=&e^{-in_c(y_a+y_b)/2}\mathcal{L}^{-1}\mathcal{J}_{\alpha_2A^2e^{2in_cy}}(\hat{C}_1^1)^*
e^{-in_c(y_a+y_b)/2}\nonumber\\ &+&e^{-in_c(y_a+y_b)/2}\mathcal{L}^{-1}
\mathcal{J}_{A^*e^{-in_cy}}\hat{C}_2^2e^{2in_c(y_a+y_b)/2},\label{eq:C312}
\end{eqnarray}
is due to the quasi-linear interactions of $\overline{u}_1$ with $\hat{C}_2^2e^{2in_c(y_a+y_b)/2}$ and
$\overline{u}_2$ with $C_1$.

The $\hat{C}_3^1$ component produces a $\mu^3$ order correction to the vorticity fluxes at wavenumber $n_c$ and
it also produces secular terms that vanish only if the amplitude $A$ satisfies
the Ginzburg-Landau (G--L) equation
\begin{equation}
c_1\partial_TA=A-c_3|A|^2A,\label{eq:GL}
\end{equation}
where
\begin{equation}
c_1=1+\frac{\epsilon_c}{4}\int\frac{\df^2{\bf k}}{2\pi}\frac{ n_c k_x^2k_{y,1} k_2^2 (k^2- n_c ^2)}
{(k^2 k^2_2 +i\beta n_c k_x k_{y,1})^2}\hat{Q}({\bf k})\ ,\label{eq:c1}
\end{equation}
\begin{eqnarray}
c_3&= &\frac{\varepsilon_c}{4}\int\frac{\df^2{\bf k}}{2\pi}\frac{ n_c  k_x^4k_2^2(k_2^2- n_c ^2)(k^2- n_c ^2)^2}{|k^2 k^2_2 +i\beta n_c k_x k_{y,1}|^2}
  \times\nonumber\\
& &\left[\frac{2k_{y,1}}{
k^2 k^2_2 +i\beta n_c k_x k_{y,1}}-\frac{k_{y,-1}}{k^2 k_{-2}^2 +i\beta n_c k_x k_{y,-1}}-\frac{k_{y,3}}{k_2^2 k_4^2
+i\beta n_c  k_x k_{y,3}}\right]\hat{Q}({\bf k})\ .\label{eq:c3_ec}
\end{eqnarray}
(Eqs. (B13) and (B16)  of
\cite{Bakas-etal-2019}). The linear term in the (G--L) equation (\ref{eq:GL}) is due to
$\hat{C}_3^{1,h}$ and drives the zonostrophic instability. The non-linear term is due to $\hat{C}_3^{1,NL}$ and
equilibrates the instability.

To obtain the eddy variance ellipse during the instability evolution as well as for the equilibrated jet, we
calculate from the covariance (\ref{eq:Cexp}) the momentum fluxes, the eddy anisotropy
and the eddy kinetic energy for a sinusoidal mean flow $\overline{u}_1=A(T)\sin(n_cy)$ as:
\begin{eqnarray}
N&=&\mu N_1+\mu^2N_2+\mu^3N_3+\cdots\nonumber\\
&=&\mu\imag\left(\hat{N}_1^1e^{in_cy}\right)+\mu^2\imag\left(\hat{N}_2^2e^{2in_cy}\right)\nonumber\\
&+&\mu^3\imag\left(\hat{N}_3^1e^{in_cy}+\hat{N}_3^3e^{3in_cy}\right)+\cdots,
\end{eqnarray}
\begin{eqnarray}
M&=&M_0+\mu M_1+\mu^2M_2+\mu^3M_3+\cdots\nonumber\\ &=&M_0+\mu\imag\left(\hat{M}_1^1e^{in_cy}\right)+\mu^2\imag\left(\hat{M}_2^0+\hat{M}_2^2e^{2in_cy}\right)
\nonumber\\ &+&
\mu^3\imag\left(\hat{M}_3^1e^{in_cy}+\hat{M}_3^3e^{3in_cy}\right)+\cdots,
\end{eqnarray}
\begin{eqnarray}
K&=&K_0+\mu K_1+\mu^2K_2+\cdots=K_0+\mu\imag\left(\hat{K}_1^1e^{in_cy}\right)\nonumber\\ &+&\mu^2\imag\left(\hat{K}_2^0+\hat{K}_2^2e^{2in_cy}\right)+\cdots
\end{eqnarray}
with $\hat{N}_j^l=e^{-iln_cy}\mathcal{N}\left(\hat{C}_j^le^{iln_c(y_a+y_b)/2}\right)$, $\hat{M}_j^l=e^{-iln_cy}\mathcal{M}\left(\hat{C}_j^le^{iln_c(y_a+y_b)/2}\right)$,  $\hat{K}_j^l=e^{-iln_cy}\mathcal{K}\left(\hat{C}_j^le^{iln_c(y_a+y_b)/2}\right)$, $M_0=\varepsilon_c\eta/2$ and $K_0=\varepsilon_c/2$.

For isotropic forcing ($\eta=0$), the tilt angle is
\begin{equation}
\theta=\theta_1+\mu\theta_2+\mu^2\theta_3+\cdots,
\end{equation}
where $\theta_1=-(1/2)\arctan\left(N_1/M_1\right)$, $\theta_2=-(M_1N_2-M_2N_1)/2(N_1^2+M_1^2)$ and the third order correction
\begin{eqnarray}
\theta_3&=&\theta_{3,2}+\theta_{3,3}\nonumber\\ &=&
-\frac{(M_1M_2+N_1N_2)(M_2N_1-M_1N_2)}{2(N_1^2+M_1^2)^2}\nonumber\\ &-&\frac{M_1N_3-M_3N_1}{2(N_1^2+M_1^2)},\label{eq:theta33}
\end{eqnarray}
contains two terms. The first ($\theta_{3,2}$) involves the second--order quantities $N_2$, $M_2$ and the second
($\theta_{3,3}$) involves the third order quantities $N_3$, $M_3$. Similarly, the anisotropy parameter is given by:
\begin{equation}
\gamma=\mu\gamma_1+\mu^2\gamma_2+\mu^3\gamma_3+\cdots,
\end{equation}
where
\begin{equation}
\gamma_1=\frac{\sqrt{N_1^2+M_1^2}}{K_0},~~\gamma_2=\frac{K_0(M_1M_2+N_1N_2)-K_1(N_1^2+M_1^2)}{K_0^2\sqrt{N_1^2+M_1^2}},
\end{equation}
and the third order correction $\gamma_3=\gamma_{3,2}+\gamma_{3,3}$ is the sum of the terms
\begin{equation}
\gamma_{3,2}=\frac{2(K_1^2-K_0K_2)(N_1^2+M_1^2)^2-2K_0K_1(M_1M_2+N_1N_2)(N_1^2+M_1^2)+K_0^2(M_2N_1-N_2M_1)^2}{2K_0^3(N_1^2+M_1^2)^{3/2}},
\end{equation}
\begin{equation}
\gamma_{3,3}=\frac{N_1N_3+M_1M_3}{K_0\sqrt{N_1^2+M_1^2}},\label{eq:gamma33}
\end{equation}
that contain second and third order quantities respectively.

The vorticity fluxes forcing the mean flow are:
\begin{equation}
\overline{v'\zeta'}=\mu f_1+\mu^2f_2+\mu^3f_3,\label{eq:vfluxexp}
\end{equation}
where
\begin{equation}
f_1=-K_0\gamma_1'\sin 2\theta_1-2K_0\gamma_1\theta_1'\cos 2\theta_1,\label{eq:f1_iso}
\end{equation}
are the fluxes at leading order that exactly overcome dissipation at the onset of instability and primed
quantities denote differentiation with respect to latitude. At the next order, the fluxes
$f_2$ drive the mean flow $\overline{u}_2$ with wavenumber $2n_c$. At the third order,
\begin{eqnarray}
f_3&=&-K_0\left(\gamma_{3,3}'-4\theta_1'\gamma_1\theta_{3,3}\right)\sin 2\theta_1\nonumber\\ &-&2K_0
\left(\theta_{3,3}'\gamma_1+\theta_1'\gamma_{3,3}+\gamma_1'
\theta_{3,3}\right)\cos\ 2\theta_1,
\end{eqnarray}
as the sum of all terms that involve second--order quantities is exactly zero. Therefore $f_3$ is a function of the third order momentum
fluxes $N_3$ and eddy anisotropy $M_3$. We observe that $|\hat{N}_3^3|\ll |\hat{N}_3^1|$ and $|\hat{M}_3^3|\ll |\hat{M}_3^1|$. As a result:
\begin{equation}
N_3\simeq \mathrm{Im}\left(\hat{N}_3^1e^{in_cy}\right)=\mathcal{N}\left(\hat{C}_3^{1,h}+\hat{C}_3^{1,NL}\right)=N_1+N_3^{NL},\label{eq:NNL}
\end{equation}
where $N_3^{NL}=N_3^{ec}+N_3^{12}=\mathcal{N}\left(\hat{C}_3^{1,ec}\right)+\mathcal{N}\left(\hat{C}_3^{1,12}\right)$. Similarly
$M_3\simeq M_1+M_3^{NL}$ with $M_3^{NL}=M_3^{ec}+M_3^{12}=\mathcal{M}\left(\hat{C}_3^{1,ec}\right)+
\mathcal{M}\left(\hat{C}_3^{1,12}\right)$. Substituting in (\ref{eq:theta33}) and (\ref{eq:gamma33})
we obtain:
\begin{eqnarray}
\theta_{3,3}&=&\theta_{3,NL}=\theta_{3,ec}+\theta_{3,12}\nonumber\\
&=&-\frac{M_1N_3^{ec}-M_3^{ec}N_1}{2(N_1^2+M_1^2)}--\frac{M_1N_3^{12}-M_3^{12}N_1}{2(N_1^2+M_1^2)},\label{eq:theta3NL}
\end{eqnarray}
\begin{eqnarray}
\gamma_{3,3}&=&\gamma_1+\gamma_{3,NL}=\gamma_1+(\gamma_{3,ec}+\gamma_{3,12})\nonumber\\
&=&\gamma_1+\frac{N_1N_3^{ec}+M_1M_3^{ec}}{K_0\sqrt{N_1^2+M_1^2}}+\frac{N_1N_3^{12}+M_1M_3^{12}}{K_0\sqrt{N_1^2+M_1^2}},\label{eq:gamma3NL}
\end{eqnarray}
where the subscripts (or superscripts) $ec$ and $12$ denote the contribution of the energy correction term
($\hat{C}_3^{1,ec}$) and the term due to the interaction of the single harmonic jet-eddy components with their double
harmonic counterparts ($\hat{C}_3^{1,12}$) to $\theta_{3,NL}$ and $\gamma_{3,NL}$.

Therefore the vorticity fluxes at third order are:
\begin{eqnarray}
f_3&=&f_1-K_0\left(\gamma_{3,NL}'-4\theta_1'\gamma_1\theta_{3,NL}\right)\sin 2\theta_1\nonumber\\
&-&2K_0\left(\theta_{3,NL}'\gamma_1+\theta_1'\gamma_{3,NL}+\gamma_1'
\theta_{3,NL}\right)\cos\ 2\theta_1.\label{eq:vorflux_final}
\end{eqnarray}
The first term corresponds to the linear term in (\ref{eq:GL}) and drives the instability, while
the other two terms correspond to the non-linear term in (\ref{eq:GL}) that equilibrate the instability. For small
jet amplitude $A$, these terms are $|A|^2$ smaller than the first term and (\ref{eq:vorflux_final})
is the leading order of
\begin{eqnarray}
f_3&\simeq& -K_0\left(\gamma_1+\gamma_{3,NL})'\sin (2\theta_1+2\theta_{3,NL}\right)\nonumber\\
&-&2K_0(\gamma_1+\gamma_{3,NL})\cos(2\theta_1+2\theta_{3,NL}).
\end{eqnarray}
Therefore $\theta_{3,NL}$ and $\gamma_{3,NL}$ are the non-linear corrections to the tilt angle and the
eddy anisotropy parameter respectively.

For anisotropic forcing ($\eta=1$), the tilt angle and the anisotropy parameter are given by:
\begin{equation}
\theta=\mu\theta_1+\mu^2\theta_2+\mu^3\theta_3+\cdots,~~~\gamma=\gamma_0+\mu\gamma_1+\mu^2\gamma_2+\cdots,
\end{equation}
where $\gamma_0=M_0/K_0$ is the anisotropy parameter for the homogeneous equilibrium, the corrections
$\theta_1=-N_1/2M_0$, $\theta_2=(M_1N_1-M_0N_2)/2M_0^2$,
\begin{eqnarray}
\gamma_1&=&\frac{K_0M_1-K_1M_0}{K_0^2},\nonumber\\ \gamma_2&=&\frac{2M_0^2(K_1^2-K_0K_2)
+K_0^2(2M_0M_2+N_1^2)}\nonumber\\ &-&\frac{2K_1M_1}{2K_0^2M_0}.
\end{eqnarray}
contain second--order quantities and the third order correction
\begin{eqnarray}
\theta_3&=&\theta_{3,2}+\theta_{3,3}\nonumber\\ &=&\frac{N_1(N_1^2-3M_1^2)+3M_0(M_2N_1+M_1N_2)}{6M_0^3}\nonumber\\
&-&\frac{N_3}{2M_0},\label{eq:theta33an}
\end{eqnarray}
has two terms containing second--order quantities ($\theta_{3,2}$) and third order quantities ($\theta_{3,3}$). The
vorticity fluxes are given by (\ref{eq:vfluxexp}), where
\begin{equation}
f_1=2\gamma_0K_0\theta_1',\label{eq:f1_noniso}
\end{equation}
balances the mean flow dissipation and
$f_2$ drives $\overline{u}_2$. The third order correction is
\begin{equation}
f_3=2\gamma_0K_0\theta_{3,3}',\label{eq:vorflux_final_an}
\end{equation}
as again the terms involving second--order quantities cancel out. Substituting (\ref{eq:NNL}) in (\ref{eq:theta33an})
we obtain $\theta_{3,3}=\theta_1+\theta_{3,NL}$, with
\begin{equation}
\theta_{3,NL}=\theta_{3,ec}+\theta_{3,12}=-\frac{N_3^{ec}}{2M_0}-\frac{N_3^{12}}{2M_0}.\label{eq:theta3NL_ani}
\end{equation}
Therefore $\theta_{3,NL}$ is
the non-linear correction to the tilt angle that brings about the equilibration of the flow according
to (\ref{eq:vorflux_final_an}). We found numerically that $N_3^{NL}=-|A|^2N_1$ and as a result
$\theta_{3,NL}=|A|^2(\pi/2-\theta_1)$.

\ifdraft\else\twocolumn\fi

\end{document}